\documentclass[aps,twocolumn,prd,showpacs,showkeys,preprintnumbers,superscriptaddress,nobibnotes,floatfix,longbibliography,nofootinbib]{revtex4-2}
\pdfoutput=1
\usepackage{amsmath}
\usepackage{amsfonts}
\usepackage{amssymb}
\usepackage{mathrsfs}
\usepackage{graphicx}
\usepackage{color}
\usepackage{hyperref}
\usepackage{placeins}
\usepackage{multirow}
\usepackage{slashed}
\usepackage{epsf}
\usepackage{cancel}

\usepackage{graphicx}
\usepackage{caption}
\usepackage{subcaption}

\usepackage{color}

\DeclareMathOperator{\ii}{i}

\usepackage{graphicx}
\usepackage{caption}
\usepackage{subcaption}

\usepackage{graphicx}
\graphicspath{{figures/}}
\usepackage{bm}
\usepackage{booktabs}
\usepackage{placeins}
\usepackage{float}
\usepackage{xcolor}
\usepackage{hyperref}
\usepackage{placeins}
\hypersetup{colorlinks=true,citecolor=blue,linkcolor=blue,urlcolor=blue}

\newcommand{\dd}{\mathrm{d}}
\renewcommand{\ii}{\mathrm{i}}
\newcommand{\ee}{\mathrm{e}}
\newcommand{\Mpl}{M_{\rm Pl}}
\newcommand{\mphi}{m_{\phi}}
\newcommand{\mA}{m_{A}}
\newcommand{\W}{W}
\newcommand{\Phii}{\Phi_i}
\newcommand{\epsi}{\epsilon_i}
\newcommand{\aosc}{a_{\rm osc}}
\newcommand{\Hosc}{H_{\rm osc}}
\newcommand{\UpsilonR}{\Upsilon}

\begin{document}

\title{Dilaton-Induced Resonant Production of Ultralight Vector Dark Matter }

\author{Imtiaz Khan}
\email{ikhanphys1993@gmail.com}
\affiliation{Department of Physics, Zhejiang Normal University, Jinhua, Zhejiang 321004, China}
\affiliation{Research Center of Astrophysics and Cosmology, Khazar University, Baku, AZ1096, 41 Mehseti Street, Azerbaijan}

\author{G. Mustafa}
\email{gmustafa3828@gmail.com}
\affiliation{Department of Physics, Zhejiang Normal University, Jinhua, Zhejiang 321004, China}

\author{Jehanzad Zafar }
\email{jehanzadzafar@stu.xjtu.edu.cn}
 \affiliation{State Key Laboratory for Mechanical Behavior of Materials, School of Materials Science and Engineering, Xi’an Jiaotong University, Xi’an 710049, P. R. China}

\author{Farruh~Atamurotov}
\email{atamurotov@yahoo.com}
\affiliation{Kimyo International University in Tashkent, Shota Rustaveli str. 156, Tashkent 100121, Uzbekistan}

 \author{Ahmadjon~Abdujabbarov}
	\email{ahmadjon@astrin.uz}
\affiliation{School of Physics, Harbin Institute of Technology, Harbin 150001, People’s Republic of China}
\affiliation{University of Tashkent for Applied Sciences, Str. Gavhar 1, Tashkent 100149, Uzbekistan}

\author{Chengxun Yuan}
   \email{yuancx@hit.edu.cn}
\affiliation{School of Physics, Harbin Institute of Technology, Harbin 150001, People’s Republic of China}

\begin{abstract}
We consider the production of ultralight vector dark matter through a dilatonic half-mass resonance in the presence of an oscillating spectator scalar. The cosmological condition is that the Floquet instability becomes efficient before the spectator contributes dominantly to the expansion. We express this condition in terms of the microscopic modulation parameter $\epsilon_i=\Phi_i/M$ and the gravitational onset fraction $r_i=\Phi_i^2/(6\Mpl^2)$. For a background with constant equation-of-state parameter $w_b$, the narrow-band result gives $\mu/H\propto a^{3w_b/2}$; the ratio therefore grows as $a^{1/2}$ during radiation domination and remains constant for matter-like expansion. Requiring delayed amplification to precede spectator domination gives the amplitude-independent bound $M/\Mpl\lesssim\sqrt6\,c_1\simeq0.31$, with $c_1\simeq1/8$ for the linear half-mass branch. An expanding-background calculation gives $a_\star<a_{\rm dom}$ for sub-Planckian $M$, whereas $M\simeq\Mpl$ delays efficient growth until after domination. Using the efficient-transfer normalization, we obtain $m_{\gamma'}\propto r_i^{-2}$, so that $m_{\gamma'}\sim10^{-20}$--$10^{-18}\,{\rm eV}$ corresponds to $r_i\sim10^{-5}$--$10^{-4}$ rather than to early spectator domination. The canonical polarization analysis shows a more infrared-localized longitudinal spectrum than the transverse spectrum, while derivative terms associated with canonical normalization change the leading Floquet exponents at order unity. St\"uckelberg and Higgsed realizations lead to different ultraviolet consistency conditions, including radial-mode decoupling and symmetry-restoration bounds. These conditions select a radiation-era, perturbative, infrared-dominated regime with a sub-Planckian kinetic scale for production.
\end{abstract}

\maketitle

\section{Introduction}

Dark matter provides one of the established indications for physics beyond the Standard Model, and ultralight bosonic fields constitute a widely studied class of candidates because of their cosmological and astrophysical consequences \cite{Hui:2017,Arias:2012,Nelson:2011,Graham:2015,Holdom:1986,Fabbrichesi:2020,Caputo:2021,Cline:2024}. Massive vector fields associated with hidden $U(1)$ gauge sectors give a simple realization of this possibility. Unlike scalar dark matter, a massive vector contains transverse and longitudinal polarization states, whose dynamics can differ during production. When the vector kinetic term varies with time, the two sectors need not be amplified in the same way. The resulting relic abundance can then depend not only on the total transferred energy, but also on the momentum distribution, polarization content, and the range in which the effective description remains applicable throughout the cosmological production and subsequent evolution.

Several mechanisms for producing vector dark matter have been considered in the literature. Hidden-photon misalignment and inflationary vacuum fluctuations provide minimal examples for generating massive vector relics \cite{Nelson:2011,Graham:2015,AlonsoAlvarez:2019cgw}. Nonadiabatic backgrounds can also produce particles exponentially through parametric resonance, as studied extensively in preheating and related early-Universe settings \cite{Traschen:1990,Shtanov:1995,Kofman:1997,Greene:1997,Amin:2014,Lozanov:2018}. For vector dark matter, resonant production has been investigated in Higgs-condensate dynamics, axion-induced tachyonic amplification, Higgsed dark-sector string networks, inflationary spectator models, and gravitational production with nonminimal couplings \cite{Dror:2018,Agrawal:2018,Co:2018,Co:2021rhi,Long:2019,Kitajima:2022strings,Kitajima:2023,Nakai:2020,Nakai:2023,Barrie:2022qni,Ema:2019yrd,Ahmed:2020fhc,Kolb:2020fwh,Cembranos:2023qph,Ozsoy:2023,Capanelli:2024runaway,Capanelli:2024gpp,Capanelli:2024}. In these scenarios, reproducing the relic abundance does not by itself specify the production history, since the momentum support, polarization dependence, background evolution, and ultraviolet completion provide additional information. We focus on a dilatonic resonance in which an oscillating scalar field $\phi$ modulates the gauge kinetic function $\W(\phi)$ of a dark vector. The resulting Floquet spectrum contains a narrow instability band whose first branch approaches the infrared near the tuned ratio $\mA/\mphi = 1/2$. This ``half-mass'' branch was identified and studied locally in Ref.~\cite{Adshead:2023}. {We use this branch as a concrete benchmark for a broader question: how local resonant growth is embedded in an expanding cosmology when a coherently oscillating scalar begins as a spectator, and how the resulting scale ordering interfaces with the ultraviolet origin of the produced field.} In this description, $\Phi/M$ sets the microscopic resonance strength, whereas $\Phi/\Mpl$ fixes the scalar contribution to the expansion. The two ratios are independent until a value of the kinetic scale $M$ is specified. The transverse Mathieu reduction identifies the local band center, whereas the longitudinal canonical system must be evaluated independently because its normalization is momentum dependent. This separation is used in the numerical analysis.

The distinction between local instability dynamics and cosmological evolution also occurs in other early-Universe settings. In dilaton-flattened and monodromic inflation, effective slopes are controlled by heavy-field or valley dynamics, while resonant axion production, gauge-field backreaction, primordial-black-hole formation, and non-Abelian dark sectors show that the microscopic production history need not be directly encoded in late-time relic observables \cite{Pirzada_2026uak,Pirzada:2026npl,Pirzada:2026sle,Pirzada:2026jml,Ijaz:2024zma,Ijaz:2023cvc,Barbon:2025wjl,Khan:2026nsz,Khan:2025ibo,Muhammad:2026gmg}. It is therefore useful to treat the local Floquet problem, the evolution of the cosmological background, and the ultraviolet origin of the vector mass as separate components of the analysis. before combining their resulting physical constraints.

We ask whether the half-mass branch can become efficient while the scalar remains energetically subdominant. For a spectator initial condition, the fractional scalar contribution at the onset of oscillations, defined by $H=\mphi$, is
\begin{equation}
 r_i = \frac{\Phii^2}{6\Mpl^2},
 \label{eq:ri_intro}
\end{equation}
where $\Phii$ is the frozen scalar amplitude at that time. This relation separates the gravitational initial condition from the microscopic resonance parameter and provides the appropriate variable for placing the local Floquet instability in an expanding cosmological background. with a separately specified dominant background component.

For the persistent narrow branch, the leading Floquet exponent may be written as $\mu \simeq c_1 (\Phi/M)\mphi$. Once a quadratic spectator oscillates coherently, its amplitude scales as $\Phi \propto a^{-3/2}$, whereas a dominant background with equation-of-state parameter $w_b$ has $H \propto a^{-3(1+w_b)/2}$. Hence
\begin{equation}
 \frac{\mu(a)}{H(a)} = \left.\frac{\mu}{H}\right|_{\aosc}
 \left(\frac{a}{\aosc}\right)^{3w_b/2}.
 \label{eq:introUpsilon}
\end{equation}
During radiation domination, or for a stiffer background, the instability becomes increasingly efficient relative to Hubble expansion, while no such enhancement occurs for matter domination. Requiring the resonance to become efficient before spectator domination gives $M/\Mpl \lesssim \sqrt6\, c_1 \simeq 0.31$ for the narrow-band coefficient $c_1 \simeq 1/8$. We check this condition directly using expanding-background trajectories for $\mu/H$, the spectator fraction, and the integrated resonance growth. so that the ordering of the characteristic scales is displayed explicitly in time.

{The scale-ordering argument uses only the quadratic-condensate scaling $\Phi\propto a^{-3/2}$ and a leading narrow-band exponent linear in $\Phi/M$. It therefore applies to the corresponding class of spectator-driven resonant production histories, with the dilatonic vector branch fixing the benchmark coefficient and polarization structure studied here.}

The relic abundance provides a separate relation. In the efficient-transfer normalization for the half-mass branch, the dark-photon mass scales as $m_{\gamma'} \propto r_i^{-2}$, linking the initial spectator fraction to the final relic mass. The interval $m_{\gamma'} \sim 10^{-20}$--$10^{-18}\,\mathrm{eV}$ then corresponds to $r_i \sim 10^{-4}$--$10^{-5}$, for which the spectator is well below the level required for early domination. We use the efficient-transfer normalization obtained in Ref.~\cite{Adshead:2023}; the pre-domination condition derived below does not depend on this abundance prescription. A causal bound on postinflationary wave dark matter may further restrict the lower end of the allowed mass interval \cite{AminMirbabayi:2024}. for the production histories considered explicitly in the present analysis.

The origin of the vector mass imposes constraints independent of cosmological embedding. In a St\"uckelberg realization, the mass does not rely on a broken-phase order parameter, so symmetry restoration and defect formation are absent, although the kinetic sector and ultraviolet cutoff must remain controlled. A Higgsed realization instead contains a radial mode and a symmetry-breaking scale. Resonantly produced vectors may then restore the symmetry, excite the radial field, or induce defect formation when the broken-phase effective description ceases to apply \cite{Holdom:1986,Fabbrichesi:2020,Long:2019,Salehian:2020jws,Redi:2022zkt,Sato:2022cwt,Kitajima:2022strings,Cyncynates:2025prl,Cyncynates:2025prd,Kitajima:2025bound}. These ultraviolet conditions are separate from the infrared Floquet criterion and are imposed independently when assessing the consistency of dark-sector realization.

The paper is organized as follows. In Sec.~\ref{sec:setup}, we formulate the polarization-resolved quadratic action for the vector field in an FLRW background. In Sec.~\ref{sec:narrow}, we obtain the small-amplitude Mathieu equation and specify the tuned half-mass branch. Section~\ref{sec:cosmo} gives the cosmological embedding, including the onset fraction, the background-dependent growth law, and the pre-domination constraint on the kinetic scale. In Sec.~\ref{sec:uv}, we discuss the ultraviolet conditions for St\"uckelberg and Higgsed realizations. Section~\ref{sec:numerics} presents numerical Floquet analysis of the transverse and longitudinal modes and compares it with analytic results. In Sec.~\ref{sec:pheno}, we use cosmological and ultraviolet conditions to obtain mass ranges and production epochs for ultralight vector dark matter. The conclusions are summarized in the final section.

{The main result of this work is the explicit separation of three quantities that are often conflated in resonant vector production: the microscopic modulation parameter $\Phi/M$, the gravitational onset fraction $r_i$, and the ultraviolet mass-generating scale. We show that the predictive power of the tuned dilatonic half-mass branch follows precisely from the simultaneous restrictions imposed by all three.}

\section{Polarization-resolved vector dynamics in an FLRW background}
\label{sec:setup}

We first formulate the polarization-resolved system used in the remainder of the analysis. The starting point is a dilatonic Proca theory in an expanding FLRW background. The longitudinal mode is canonically normalized explicitly, since its normalization determines part of the momentum dependence of the resonance. Related canonical treatments of massive vector fields have been used in analyses of inflationary fluctuation production, Higgs-condensate resonance, gravitational production in cosmological backgrounds, nonminimal curvature couplings, and dilatonic parametric amplification \cite{Graham:2015,Dror:2018,Ema:2019yrd,Ahmed:2020fhc,Kolb:2020fwh,AlonsoAlvarez:2019cgw,Ozsoy:2023,Capanelli:2024gpp,Adshead:2023}. We use the same field-theoretic framework while keeping the local resonance variables distinct from the quantities controlling the cosmological evolution of the spectator. We consider a spectator scalar field $\phi$ with a quadratic potential and a massive vector field $A_\mu$ whose kinetic term depends on the scalar through $\W(\phi)$. The action is
\begin{equation}
\begin{aligned}
 S = \int \dd^4x \sqrt{-g}\Bigg[&\frac{\Mpl^2}{2}R + \frac12 \, \partial_\mu \phi\, \partial^\mu \phi - \frac12 \, \mphi^2 \phi^2 \\
 &- \frac14 \, \W(\phi) F_{\mu\nu}F^{\mu\nu} + \frac12 \, \mA^2 A_\mu A^\mu\Bigg].
\end{aligned}
 \label{eq:action}
\end{equation}
Here $g$ is the metric determinant, $R$ is the Ricci scalar, and $F_{\mu\nu} \equiv \partial_\mu A_\nu - \partial_\nu A_\mu$ denotes the vector field-strength tensor. The function $\W(\phi)$ specifies the gauge-kinetic normalization. For a homogeneous scalar background, $\phi=\bar{\phi}(\tau)$, it is a function of time only; below we write $\W(\bar{\phi}(\tau)) \rightarrow \W$. We take the background metric to be spatially flat Friedmann-Lema\^itre-Robertson-Walker spacetime in conformal time $\tau$,
$ \dd s^2 = a(\tau)^2(\dd \tau^2 - \dd \bm{x}^2),$
where $a(\tau)$ denotes the scale factor. In the parameter region considered here, the scalar is subdominant until the Hubble rate approaches $H \simeq \mphi$. Its homogeneous solution is therefore overdamped for $H \gg \mphi$ and subsequently enters coherent oscillations in the quadratic potential for $H \lesssim \mphi$ \cite{Allahverdi:2010,Kofman:1997,Amin:2014,Lozanov:2018}. This separation allows the local vector instability to be analyzed on a prescribed cosmological background before the backreaction conditions are imposed. It also keeps the gravitational and microscopic expansion parameters explicitly distinct throughout the derivation below.

We separate the physical polarizations by decomposing the spatial vector field in Fourier space. For comoving momentum $\bm{k}$ and $\hat{\bm{k}} \equiv \bm{k}/k$, we write
\begin{equation}
 A_i(\tau,\bm{k}) = \sum_{\lambda=\pm} \epsilon_i^{(\lambda)}(\hat{\bm{k}}) A_\lambda(\tau,k) + \hat{k}_i A_L(\tau,k),
\end{equation}
where $\epsilon_i^{(\lambda)}$ are the transverse polarization vectors and satisfy $\hat{\bm{k}}\cdot \bm{\epsilon}^{(\lambda)}=0$, $A_\lambda$ denote the transverse amplitudes, and $A_L$ is the longitudinal mode. Expanding Eq.~\eqref{eq:action} to quadratic order in the vector fluctuations, transforming to momentum space, and using the homogeneity of $\W$, we obtain
\begin{equation}
\begin{aligned}
 S_A^{(2)} = \frac12 \int \frac{\dd^3k\,\dd\tau}{(2\pi)^3}
 \Big[& \W\, |A_i'-\ii k_i A_0|^2 \\
 &- \W\,(k^2\delta_{ij}-k_i k_j)A_i A_j^* \\
 &+ a^2\mA^2 |A_0|^2 - a^2\mA^2 A_iA_i^*\Big].
\end{aligned}
 \label{eq:vectorquad}
\end{equation}
The temporal component $A_0$ has no independent dynamics and acts as a constraint variable. The transverse and longitudinal quadratic systems follow by projecting Eq.~\eqref{eq:vectorquad} onto the corresponding polarization subspaces. This form also makes the separate treatment of the two canonical sectors transparent.

\subsection{Transverse sector}

For transverse modes, $k_i \epsilon_i^{(\lambda)}=0$ removes the mixing with the nondynamical field $A_0$. The conformal invariance of the Maxwell kinetic term also prevents an additional scale-factor factor from appearing in the kinetic contribution. The quadratic action is therefore
\begin{equation}
\begin{aligned}
 S_T = \frac12 \sum_{\lambda=\pm}
 \int \frac{\dd^3k\,\dd\tau}{(2\pi)^3}
 \Big[& \W\, |A_\lambda'|^2 \\
 &- \left(\W k^2 + a^2\mA^2\right)|A_\lambda|^2\Big].
\end{aligned}
 \label{eq:ST}
\end{equation}
With the canonical transverse variable
\begin{equation}
 v_\lambda \equiv \sqrt{\W}\, A_\lambda,
\end{equation}
the mode equation takes the oscillator form
\begin{equation}
 v_\lambda'' + \Omega_T^2(\tau,k) v_\lambda = 0,
 \qquad
 \Omega_T^2 = k^2 + \frac{a^2\mA^2}{\W} - \frac{(\sqrt{\W})''}{\sqrt{\W}}.
 \label{eq:transverseeq}
\end{equation}
The last term is generated by canonical normalization. It is of the same perturbative order as the modulation of $a^2\mA^2/\W$ and must therefore be retained when the tuned instability is treated consistently in the small-amplitude expansion.

\subsection{Longitudinal sector}

The longitudinal sector is different because the scalar-like mode $A_L$ mixes with the nondynamical temporal component $A_0$. Projecting Eq.~\eqref{eq:vectorquad} onto $A_i=\hat{k}_iA_L$ and varying with respect to $A_0$ gives
\begin{equation}
 A_0(\tau,k) = \frac{\ii k\, \W}{k^2\W + a^2\mA^2}\, A_L'(\tau,k).
 \label{eq:A0constraint}
\end{equation}
Substitution of this algebraic relation into the quadratic action eliminates the auxiliary field and gives
\begin{equation}
 S_L = \frac12 \int \frac{\dd^3k\,\dd\tau}{(2\pi)^3}\left[ \frac{\W a^2\mA^2}{k^2\W + a^2\mA^2} |A_L'|^2 - a^2\mA^2 |A_L|^2\right].
 \label{eq:SL}
\end{equation}
We define the canonical normalization by
\begin{equation}
 z_L^2(\tau,k) \equiv \frac{\W a^2\mA^2}{k^2\W + a^2\mA^2},
 \qquad v_L \equiv z_L A_L,
 \label{eq:zLdef}
\end{equation}
which yields the longitudinal oscillator equation
\begin{equation}
 v_L'' + \Omega_L^2(\tau,k) v_L = 0,
 \qquad
 \Omega_L^2 = k^2 + \frac{a^2\mA^2}{\W} - \frac{z_L''}{z_L}.
 \label{eq:longitudinaleq}
\end{equation}
The momentum dependence of the normalization factor is retained explicitly in the calculations below.

Equations~\eqref{eq:transverseeq} and \eqref{eq:longitudinaleq} display the source of the polarization dependence. Both sectors contain the same bare frequency, $k^2 + a^2\mA^2/\W$, but the canonical terms generated by the two field redefinitions are different. The longitudinal term appears only after the temporal constraint has been eliminated and therefore has a separate momentum dependence. The longitudinal response consequently cannot be obtained from the transverse result by a multiplicity factor. The difference is largest near $k^2\W \sim a^2\mA^2$, where the normalization in Eq.~\eqref{eq:zLdef} varies rapidly and increases the infrared sensitivity of the longitudinal mode. This distinction will be retained in both the analytic and numerical analyses.

\section{Small-amplitude expansion and the tuned branch}
\label{sec:narrow}

We next consider the perturbative small-amplitude limit of the tuned half-mass branch. The calculation uses the standard Floquet treatment of periodic instabilities \cite{Traschen:1990,Shtanov:1995,Kofman:1997,Greene:1997,Amin:2014,Lozanov:2018} and follows the same general approach used for vector production from oscillating masses, kinetic functions, and axionlike couplings \cite{Dror:2018,Co:2018,Co:2021rhi,Kitajima:2023,Nakai:2023,Barrie:2022qni,Adshead:2023}. Here the expansion is used to identify the canonical frequency modulation governing the local resonance and to obtain the quantities needed for the cosmological consistency analysis below. This step requires no assumption about later abundance normalization or the ultraviolet origin of vector mass.

We take the gauge-kinetic function to have the exponential form
\begin{equation}
 \W(\phi) = \exp\!\left(\frac{\phi}{M}\right),
 \label{eq:Wexp}
\end{equation}
and write the homogeneous spectator solution as
\begin{equation}
 \bar\phi(t) = \Phi(t)\cos(\mphi t),
 \qquad
 \epsilon(t) \equiv \frac{\Phi(t)}{M}.
 \label{eq:phiosc}
\end{equation}
The slowly varying envelope is denoted by $\Phi(t)$, and $\epsilon(t)$ is the dimensionless parameter controlling the small-amplitude expansion. In the quasi-adiabatic regime $H \ll \mphi$, the scale factor changes negligibly during a single scalar oscillation. The local resonance problem may then be treated in an approximately Minkowski background, with cosmological redshifting included on longer time scales.

To first order in $\epsilon$, the kinetic factors are
\begin{equation}
\begin{aligned}
 \W^{-1} &= 1 - \epsilon \cos(\mphi t) + {\cal O}(\epsilon^2), \\
 \frac{\ddot{\sqrt{\W}}}{\sqrt{\W}} &= -\frac12 \, \epsilon \, \mphi^2 \cos(\mphi t) + {\cal O}(\epsilon^2), \\
 -\frac{\ddot{\sqrt{\W}}}{\sqrt{\W}} &= \frac12 \, \epsilon \, \mphi^2 \cos(\mphi t) + {\cal O}(\epsilon^2).
\end{aligned}
\end{equation}
Substituting these expressions in the transverse canonical equation~\eqref{eq:transverseeq} gives
\begin{equation}
 \ddot v_k + \left[\omega_{k0}^2 + \epsilon\,\Delta_T^2\cos(\mphi t) + {\cal O}(\epsilon^2)\right] v_k = 0,
 \label{eq:TransExpand}
\end{equation}
with the unperturbed frequency and the linear modulation coefficient
\begin{equation}
 \omega_{k0}^2 = k^2 + \mA^2,
 \qquad
 \Delta_T^2 = \frac12\mphi^2 - \mA^2.
 \label{eq:omegadef}
\end{equation}
Thus, both the mass modulation and the derivative contribution enter the first-order periodic driving term.

Introducing the dimensionless time coordinate
$
 z = {\mphi t}/{2},
$
transforms Eq.~\eqref{eq:TransExpand} into the Mathieu form
\begin{equation}
 \frac{\dd^2 v_k}{\dd z^2} + \left[A_k - 2 q_T \cos(2z)\right]v_k = 0.
\end{equation}
The corresponding parameters are
\begin{equation}
\begin{aligned}
 A_k &= \frac{4(k^2+\mA^2)}{\mphi^2} + {\cal O}(\epsilon^2), \\
 q_T &= -2\epsilon\left(\frac12 - \frac{\mA^2}{\mphi^2}\right)
 + {\cal O}(\epsilon^2).
\end{aligned}
 \label{eq:MathieuPars}
\end{equation}
These expressions determine the position and width of the narrow instability bands at leading order.

The standard Mathieu instability bands then determine the resonance momenta. The first narrow band is centered at $A_k = 1$, which gives
\begin{equation}
 k_{\rm res}^2 = \frac{\mphi^2}{4} - \mA^2 + {\cal O}(\epsilon^2).
 \label{eq:bandcenter}
\end{equation}
This expression fixes the momentum of the leading transverse instability. Its center reaches the infrared at zero momentum, giving the tuned relation
\begin{equation}
 \mA = \frac{\mphi}{2}.
 \label{eq:halfmass}
\end{equation}
This is the half-mass condition used in the remainder of the analysis.

Equation~\eqref{eq:halfmass} specifies the tuned half-mass branch. Here the resonance is selected by the kinematic alignment of the first narrow band with the infrared, rather than by a large coupling or a broad-resonance regime. Equation~\eqref{eq:bandcenter} therefore gives the location of the leading transverse instability in the perturbative limit $\epsilon \ll 1$. The complete vector system, however, remains polarization dependent. In particular, the longitudinal equation contains the independent canonical term $z_L''/z_L$ in Eq.~\eqref{eq:longitudinaleq}, which changes its effective periodic frequency. A consistent treatment at the half-mass point must therefore retain the transverse and longitudinal sectors separately \cite{Dror:2018,Graham:2015,Nakai:2020,Adshead:2023,Kitajima:2023}. {The half-mass system provides the benchmark local Floquet realization used below. The cosmological embedding developed in Sec.~\ref{sec:cosmo} is controlled by the background scaling of the spectator amplitude and the local Floquet exponent, while the polarization and detuning analyses retain the full vector dynamics of the benchmark model.} The transverse Mathieu reduction identifies the band center, whereas the longitudinal system must be evaluated independently because its normalization is momentum dependent. This separation is used below.

\section{Cosmological embedding}
\label{sec:cosmo}
{The local narrow-resonance expansion is controlled by $\epsilon=\Phi/M$, whereas the spectator fraction enters the Friedmann equation. These quantities play different roles: $\epsilon$ measures the microscopic modulation, while $r_i$ measures the gravitational importance of the scalar background.}

\subsection{Onset fraction}

Before coherent oscillations begin, $H \gg \mphi$ and the spectator is overdamped, so its displacement remains approximately constant. Its energy density is then
\begin{equation}
 \rho_\phi \simeq \frac12 \mphi^2 \Phii^2,
 \qquad (H \gg \mphi).
 \label{eq:rhophi}
\end{equation}
Assuming that the spectator is subdominant in this regime, coherent oscillations start when the Hubble rate reaches the scalar mass,
\begin{equation}
 H(\aosc) = \Hosc = \mphi.
\end{equation}
The Friedmann equation gives the background energy density at this time as
\begin{equation}
 \rho_b(\aosc) = 3\Mpl^2\mphi^2.
 \label{eq:rhobosc}
\end{equation}
The spectator fraction at the onset of oscillations is consequently
\begin{equation}
 r_i \equiv \left.\frac{\rho_\phi}{\rho_b}\right|_{H=\mphi} = \frac{\Phii^2}{6\Mpl^2}.
 \label{eq:ri}
\end{equation}
This quantity is defined before the later redshifting of either component becomes relevant. These definitions fix the normalization used for the subsequent background evolution.

This relation depends only on the background at the onset of oscillations and not on the subsequent expansion history. It is therefore the appropriate measure of the initial spectator contribution to the total energy density. The spectator is already dynamically relevant at onset only if
\begin{equation}
 r_i \gtrsim 1
 \qquad \Longleftrightarrow \qquad
 \Phii \gtrsim \sqrt{6}\,\Mpl.
 \label{eq:relevance}
\end{equation}
This condition constrains the gravitational ratio $\Phii/\Mpl$ alone. It does not constrain the independent microscopic ratio $\Phii/M$, which determines the strength of the local resonance.

\subsection{General-$w_b$ evolution after onset}

After the scalar starts coherent oscillations in a quadratic potential, its averaged energy density redshifts as pressureless matter \cite{Kofman:1997,Amin:2014,Lozanov:2018},
\begin{equation}
 \rho_\phi(a) = \rho_\phi(\aosc)\left(\frac{a}{\aosc}\right)^{-3}.
\end{equation}
We take the dominant cosmological component to have a constant equation-of-state parameter $w_b$. Its energy density and the corresponding Hubble rate then scale as
\begin{equation}
\begin{aligned}
 \rho_b(a) &= \rho_b(\aosc)\left(\frac{a}{\aosc}\right)^{-3(1+w_b)}, \\
 H(a) &= \mphi \left(\frac{a}{\aosc}\right)^{-\frac32(1+w_b)}.
\end{aligned}
 \label{eq:Hscaling}
\end{equation}
The ratio of the spectator density to the background density therefore evolves according to
\begin{equation}
 r(a) = r_i\left(\frac{a}{\aosc}\right)^{3w_b}.
 \label{eq:rofA}
\end{equation}
This expression will be used to compare the resonance time with the onset of spectator domination.

This relation shows that the post-onset evolution is set by the equation of state of the dominant background. For $w_b>0$, the spectator fraction grows and reaches unity at
\begin{equation}
 \frac{a_{\rm dom}}{\aosc} = r_i^{-1/(3w_b)},
 \qquad
 \frac{H_{\rm dom}}{\mphi} = r_i^{\frac{1+w_b}{2w_b}}.
 \label{eq:domination}
\end{equation}
For radiation domination, $w_b=1/3$, one recovers $r(a)\propto a$, so the spectator becomes progressively more important as the Universe expands. The matter-dominated case, $w_b=0$, is different: $r(a)$ remains constant, and a spectator that is subdominant at onset cannot overtake the background unless the equation of state changes at a later time.

\subsection{Growth against Hubble dilution}

For the tuned half-mass branch, the cosmological evolution can be expressed directly in terms of the local narrow-band exponent. In the perturbative regime,
\begin{equation}
 \mu(a) = c_1\,\epsilon(a)\,\mphi + {\cal O}(\epsilon^2),
 \label{eq:mudef}
\end{equation}
where $c_1$ is the dimensionless coefficient appropriate to the chosen point in the instability band. For oscillations in a quadratic potential, $\Phi(a)\propto a^{-3/2}$, and hence
\begin{equation}
 \epsilon(a) = \epsi\left(\frac{a}{\aosc}\right)^{-3/2}.
 \label{eq:epsscaling}
\end{equation}
Using Eqs.~\eqref{eq:Hscaling}, \eqref{eq:mudef}, and \eqref{eq:epsscaling}, the ratio of the local Floquet rate to the Hubble rate is
\begin{equation}
 \UpsilonR(a) \equiv \frac{\mu(a)}{H(a)}
 = \UpsilonR_{\rm osc}\left(\frac{a}{\aosc}\right)^{3w_b/2},
 \qquad
 \UpsilonR_{\rm osc} = c_1\epsi.
 \label{eq:UpsilonScaling}
\end{equation}
This form separates the local instability coefficient from the subsequent background redshifting.

Equation~\eqref{eq:UpsilonScaling} gives the competition between Floquet growth and Hubble expansion. For radiation domination, $w_b=1/3$, one has $\mu/H\propto a^{1/2}$, so the relative efficiency of the resonance increases with time. For matter domination, $w_b=0$, the ratio is constant, while for $w_b>1/3$ it grows more rapidly. The tuned branch is therefore favored by backgrounds that redshift faster than pressureless matter, provided the spectator remains subdominant during the relevant interval.

{Equation~\eqref{eq:UpsilonScaling} follows from three ingredients: quadratic-condensate redshifting, the background equation of state, and a leading Floquet exponent linear in the oscillation amplitude. Any spectator-driven narrow resonance satisfying these ingredients obeys the same scale ordering, with model dependence entering through $c_1$ and the instability spectrum.}

For $w_b>0$ and an initially inefficient resonance, $\UpsilonR_{\rm osc}<1$, we define $a_\star$ by $\UpsilonR(a_\star)=1$. Equation~\eqref{eq:UpsilonScaling} gives
\begin{equation}
 \frac{a_\star}{\aosc} = \UpsilonR_{\rm osc}^{-2/(3w_b)}.
 \label{eq:astar}
\end{equation}
Comparing this scale with Eq.~\eqref{eq:domination}, the resonance becomes efficient before spectator domination when
\begin{equation}
 c_1^2 \epsi^2 \gtrsim r_i.
 \label{eq:beforeDom}
\end{equation}
Using the onset relation in Eq.~\eqref{eq:ri}, the same condition can be written as
\begin{equation}
 c_1 \gtrsim \frac{M}{\sqrt{6}\,\Mpl}.
 \label{eq:criterionM}
\end{equation}
The dependence on the initial oscillation amplitude cancels in this comparison.

Equation~\eqref{eq:criterionM} is the dynamical condition for placing the tuned branch in an expanding spectator background. It differs from abundance matching because it follows only from requiring the resonance to become efficient before the scalar changes the background expansion. Equivalently,
\begin{equation}
 \frac{M}{\Mpl} \lesssim \sqrt{6}\,c_1
 \simeq 0.31\left(\frac{c_1}{1/8}\right),
 \label{eq:Mupper}
\end{equation}
where the small-amplitude Floquet calculation gives $c_1\simeq0.125$ for both canonical polarizations at the center of the first tuned band. This value refers to the linear narrow-resonance regime; away from the band center, Eq.~\eqref{eq:Mupper} applies after replacing $c_1$ by the corresponding local Floquet coefficient. The physical content of the bound is straightforward. At fixed gravitational amplitude, increasing the kinetic scale reduces the microscopic modulation and delays the onset of efficient resonance relative to spectator domination. The condition therefore disfavors Planckian and super-Planckian kinetic scales for the tuned half-mass branch, including parameter choices considered in high-scale dilatonic or nonminimal vector constructions \cite{Fabbrichesi:2020,Ozsoy:2023,Capanelli:2024runaway,Capanelli:2024gpp}. For the benchmark $M=10^{17}\,{\rm GeV}\simeq0.041\Mpl$ used below, Eq.~\eqref{eq:Mupper} is satisfied. Since $\epsilon_i$ cancels, the result constrains the kinetic scale directly. It is nevertheless a necessary rather than sufficient condition: the initial amplitude must also allow enough integrated growth before the oscillation envelope is redshifted to smaller values. Thus, the condition does not state that resonance is guaranteed; it only fixes whether the local growth rate can overtake Hubble dilution before the background hierarchy changes.

\subsection{Expanding-background validation of the bound}

The pre-domination condition can be checked by following the evolution in an expanding background. For radiation domination, Eqs.~\eqref{eq:UpsilonScaling} and \eqref{eq:domination} become
\begin{equation}
 \UpsilonR(a)=c_1\epsi\left(\frac{a}{\aosc}\right)^{1/2},
 \qquad
 r_\phi(a)=r_i\left(\frac{a}{\aosc}\right),
 \label{eq:trajUR}
\end{equation}
and the accumulated Floquet growth is
\begin{equation}
 {\cal N}(a)\equiv \int_{\aosc}^{a}\frac{\mu}{H}\,\dd\ln \tilde a
 =2c_1\epsi\left[\left(\frac{a}{\aosc}\right)^{1/2}-1\right].
 \label{eq:cumgrowth}
\end{equation}
We use ${\cal N}\sim1$ as an estimate of order-one local amplification. At fixed $\epsilon_i$, the resulting trajectories can be compared for sub-Planckian and Planckian kinetic scales. For the sub-Planckian benchmark, both ${\cal N}=1$ and $\UpsilonR=1$ are reached before the spectator fraction becomes of order unity. For the Planckian benchmark, spectator domination occurs first and the tuned linear branch does not become efficient during radiation domination. The trajectory therefore gives a direct time-domain interpretation of Eq.~\eqref{eq:Mupper}. The calculation is restricted to the linear Floquet regime; depletion of the scalar condensate and backreaction from the produced vectors are not included. This comparison also distinguishes an instantaneous efficiency criterion from the accumulated growth criterion, which need not be reached at exactly the same scale factor along a trajectory.

\begin{figure*}[tbp]
\centering
\includegraphics[width=1.0\textwidth]{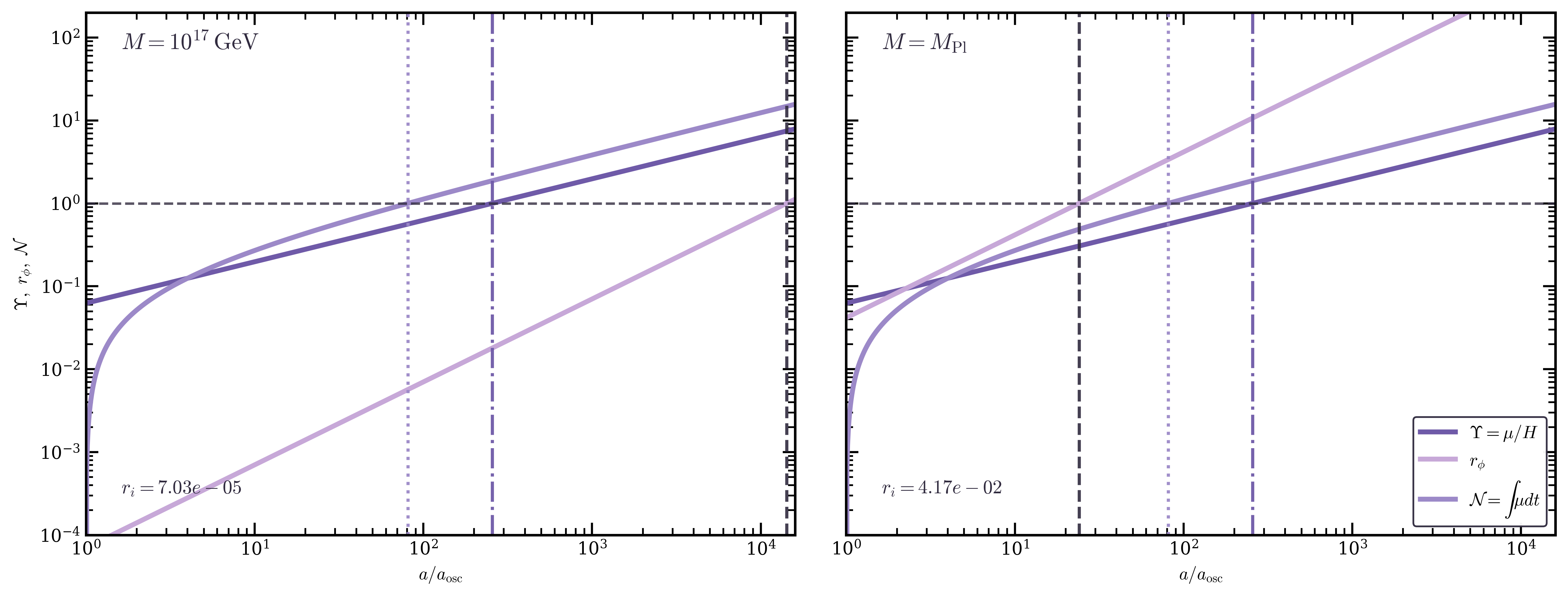}
\caption{Evolution of the pre-domination quantities during radiation domination for $c_1=1/8$ and $\epsilon_i=0.5$. We show $\Upsilon=\mu/H$, the spectator fraction $r_\phi$, and the accumulated growth ${\cal N}=\int\mu dt$ as functions of $a/a_{\rm osc}$. Left: for $M=10^{17}\,{\rm GeV}\simeq0.041\Mpl$, the conditions ${\cal N}=1$ and $\Upsilon=1$ are reached before spectator domination. Right: for $M=\Mpl$, spectator domination occurs before the delayed resonance becomes efficient. The two trajectories illustrate the ordering associated with the bound $M/\Mpl\lesssim\sqrt6\,c_1$ in Eq.~\eqref{eq:Mupper}.}
\label{fig:trajectory}
\end{figure*}

\subsection{Consistency map}

Combining Eq.~\eqref{eq:ri} with $\epsi = \Phii/M$ gives the exact relation
\begin{equation}
 r_i = \frac{\epsi^2}{6}\left(\frac{M}{\Mpl}\right)^2.
 \label{eq:exactmap}
\end{equation}
This equation maps the microscopic oscillation parameter to the gravitational onset fraction. Considered alone, Eq.~\eqref{eq:exactmap} shows that maintaining a perturbative regime with $\epsi\ll1$ while approaching gravitational relevance would require $M\gg\Mpl$. Equation~\eqref{eq:Mupper} supplies the independent dynamical condition that the resonance become efficient before spectator domination, namely $M\lesssim0.31\Mpl$ for the measured tuned-branch coefficient. In the $(\epsilon_i,M/\Mpl)$ plane, contours of constant $r_i$ satisfy
$
 M/\Mpl=\sqrt{6r_i}/\epsilon_i,
$
whereas Eq.~\eqref{eq:Mupper} gives a horizontal upper boundary. The two relations show that the perturbative tuned branch operates away from early spectator domination. They also make explicit that the microscopic resonance parameter and the gravitational backreaction parameter are not interchangeable. The map is useful because the same microscopic modulation can correspond to different gravitational initial conditions when the kinetic scale is varied, even though the local small-amplitude Floquet problem is unchanged.

\section{Ultraviolet consistency and the origin of the vector mass}
\label{sec:uv}

The preceding section constrains the cosmological evolution of the spectator and the time at which the tuned branch can become efficient. These conditions do not by themselves establish the ultraviolet consistency of the vector theory. The origin of the vector mass introduces additional requirements that must be imposed separately. In the present setup, three questions are relevant: whether the Proca theory is valid at the production scales, whether the time-dependent kinetic function drives the gauge sector outside the perturbative regime, and, for a Higgsed realization, whether the produced vectors destabilize the broken phase or lead to defect formation \cite{Holdom:1986,Fabbrichesi:2020,Ozsoy:2023,Capanelli:2024runaway,Capanelli:2024gpp,Long:2019,Salehian:2020jws,Redi:2022zkt,Sato:2022cwt,Kitajima:2022strings,Cyncynates:2025prl,Cyncynates:2025prd,Kitajima:2025bound}. These requirements are independent of the scale-ordering argument and provide separate constraints on the same resonance branch. We therefore discuss the kinetic sector and mass completion separately below. The purpose is to keep infrared dynamics separate from assumptions about high-energy completion during production.

\subsection{Control of the kinetic modulation}

Canonical normalization also affects interactions with charged matter because the effective gauge coupling depends on the kinetic function. If the vector couples to a dark-sector current $J^\mu$ through $gJ^\mu A_\mu$, the canonical coupling is
\begin{equation}
 g_{\rm eff}(\phi) = \frac{g}{\sqrt{\W(\phi)}}.
 \label{eq:geff}
\end{equation}
For the exponential form in Eq.~\eqref{eq:Wexp}, the oscillating spectator gives
\begin{equation}
 g_{\rm eff}(t)= g\,\exp\!\left[-\frac{\bar\phi(t)}{2M}\right],
 \qquad
 g_{\rm eff}^{\rm max}= g\,\ee^{\epsilon_i/2},
 \label{eq:geffmax}
\end{equation}
where the maximum occurs near the minimum of $\W$. Perturbative control throughout an oscillation therefore requires
\begin{equation}
 g\,\ee^{\epsilon_i/2} \lesssim 4\pi.
 \label{eq:pertcontrol}
\end{equation}
This condition concerns the gauge-sector coupling and is separate from the Floquet efficiency criterion.

For the perturbative tuned branch with $\epsilon_i\lesssim1$, this constraint is usually mild when the dark gauge sector is weakly or moderately coupled. It nevertheless represents an independent ultraviolet condition and should be imposed separately from the resonance calculation \cite{Fabbrichesi:2020,Ozsoy:2023,Capanelli:2024runaway,Capanelli:2024gpp,Adshead:2023}.

\subsection{St\"uckelberg completion}

A gauge-invariant vector mass can be introduced through the St\"uckelberg Lagrangian
\begin{equation}
 \mathcal{L}_{\rm St} = -\frac14 \W(\phi) F_{\mu\nu}F^{\mu\nu} - \frac12 \left(\partial_\mu \sigma - m_S A_\mu\right)^2.
 \label{eq:stuck}
\end{equation}
In unitary gauge, Eq.~\eqref{eq:stuck} reduces to the Proca theory with $\mA=m_S$. Since the mass does not originate from spontaneous symmetry breaking, there is no broken-phase order parameter that can be restored during production, and the corresponding defect constraints of a Higgsed theory do not arise. The remaining issue is the validity of the St\"uckelberg effective description over the full production interval. The characteristic resonance scales must stay below the ultraviolet cutoff and below charged thresholds that could generate appreciable kinetic mixing or higher-dimensional interactions \cite{Nelson:2011,Arias:2012,Fabbrichesi:2020,Ozsoy:2023,Capanelli:2024runaway,Capanelli:2024gpp}. When these conditions are satisfied, the infrared tuned branch may be described consistently by the Proca effective theory. This is an effective-theory requirement and does not alter the local half-mass resonance equations.

\subsection{Higgsed completion and symmetry restoration}

For a Higgsed realization, the vector mass follows from spontaneous breaking of the dark gauge symmetry by a charged scalar $\Sigma$,
\begin{equation}
\begin{aligned}
 \mathcal{L}_{\rm H} ={}& |D_\mu\Sigma|^2 - \frac{\lambda_\Sigma}{4}\left(|\Sigma|^2-v^2\right)^2 \\
 &- \frac14 \, \W(\phi)F_{\mu\nu}F^{\mu\nu},
 \qquad D_\mu = \partial_\mu + \ii g A_\mu,
\end{aligned}
 \label{eq:higgsL}
\end{equation}
with $\mA=gv$ in the broken phase. Here $v$ is the symmetry-breaking vacuum expectation value. Writing
$
 \Sigma=(v+h)\ee^{\ii\theta/v}/\sqrt{2},
$
separates the radial fluctuation $h$ from the Goldstone mode $\theta$, and the radial mass is $m_h^2=\lambda_\Sigma v^2$. These additional degrees of freedom must remain consistent with the resonant production history.

The produced vector field affects the Higgs sector through the covariant derivative and shifts the effective radial mass,
\begin{equation}
 m_{h,\rm eff}^2 \simeq -\lambda_\Sigma v^2 + g^2\langle A_\mu A^\mu\rangle + \cdots.
 \label{eq:mheff}
\end{equation}
When the gauge-field contribution becomes comparable to the symmetry-breaking term, the broken vacuum can be destabilized. A conservative condition for symmetry restoration is
\begin{equation}
 g^2 \langle A_i A_i\rangle \gtrsim \lambda_\Sigma v^2.
 \label{eq:restore1}
\end{equation}
For nonrelativistic produced vectors,
$
 \rho_A \simeq 1/2 \mA^2\langle A_i A_i\rangle,
$
and Eq.~\eqref{eq:restore1} becomes
\begin{equation}
 \rho_A \gtrsim \frac12 \lambda_\Sigma v^4 = \frac{m_h^2 v^2}{2}.
 \label{eq:restore2}
\end{equation}
This estimate identifies the energy-density scale at which the Higgsed description may cease to remain in the broken phase.

This relation gives a first ultraviolet condition for a Higgsed completion: the vector energy density should remain below the scale at which the broken phase is destabilized. For a benchmark that reproduces the observed dark-matter abundance, this condition can be written at the production epoch as
\begin{equation}
\begin{aligned}
 m_h v &\gtrsim \sqrt{2\rho_{\rm DM,0}}\,(1+z_\star)^{3/2} \\
 &\simeq 4.0\times10^2\,{\rm eV}^2
 \left(\frac{1+z_\star}{2\times10^5}\right)^{3/2} .
\end{aligned}
 \label{eq:higgsbench}
\end{equation}
where $\rho_{\rm DM,0}\simeq10^{-11}\,{\rm eV}^4$ is the present dark-matter density. This requirement does not depend on the spectator onset fraction and is imposed independently of Eq.~\eqref{eq:Mupper}. It is generally weak for a high-scale Higgs sector but can become relevant for dark sectors with symmetry-breaking scales in the eV--keV range.

A second condition is the decoupling of the radial mode. The Proca description is reliable only when the radial Higgs excitation is adiabatically heavy compared with the scales entering the resonance,
\begin{equation}
 m_h \gg \mphi,
 \qquad
 m_h \gg \frac{k_{\rm peak}}{a},
 \label{eq:radialdecouple}
\end{equation}
so that the broken-phase radial dynamics can be integrated out. This requirement remains relevant for the tuned branch because its instability is concentrated in the infrared but has a finite momentum width \cite{Dror:2018,Long:2019,Salehian:2020jws,Redi:2022zkt,Sato:2022cwt,Kitajima:2022strings,Kitajima:2023}.

A third condition concerns topological defects. If Eq.~\eqref{eq:restore1} is satisfied in the direction of restoration, the system can return to the symmetric phase. Subsequent symmetry breaking may then produce a cosmic-string network with tension
\begin{equation}
 \mu_{\rm str} \sim \pi v^2 \ln\!\left(\frac{m_h}{\mA}\right),
 \label{eq:stringtension}
\end{equation}
up to the usual logarithmic dependence on the scalar and vector core scales. In this situation the final vector abundance is not determined by homogeneous resonance alone, because string emission and loop evolution also contribute to the production history \cite{Long:2019,Salehian:2020jws,Redi:2022zkt,Sato:2022cwt,Kitajima:2022strings,Cyncynates:2025prl,Cyncynates:2025prd}.

Dark-Higgs backreaction can restrict the parameter space of minimal dark-photon dark-matter models through symmetry restoration and subsequent defect formation \cite{Mirizzi:2009,Redi:2022zkt,Sato:2022cwt,Cyncynates:2025prl,Cyncynates:2025prd,Kitajima:2025bound}. These constraints are independent of the gravitational condition in Eq.~\eqref{eq:ri}. The latter is fixed by the initial spectator amplitude and the cosmological background, whereas the Higgs-sector constraints depend on the produced vector density and on the ultraviolet parameters $v$, $m_h$, and $g$. A Higgsed realization must therefore satisfy the cosmological embedding condition together with Eq.~\eqref{eq:pertcontrol}, the radial-decoupling condition in Eq.~\eqref{eq:radialdecouple}, and the restoration bound in Eq.~\eqref{eq:restore2}. For a St\"uckelberg mass, the defect constraints are absent, while the kinetic-sector and cutoff requirements remain. The tuned branch describes the infrared production dynamics; the ultraviolet realization determines whether this dynamics remains within a controlled effective theory. The distinction is useful because satisfying the cosmological bound does not by itself guarantee that the assumed low-energy vector description remains valid during efficient production.

\section{Instability charts, polarization structure, and detuning sensitivity}
\label{sec:numerics}

The analytic results above imply three features of the tuned half-mass branch. First, for $\mA/\mphi=1/2$, the leading instability should remain concentrated near the infrared. Second, the transverse and longitudinal modes should exhibit different growth patterns rather than reducing to a single scalar-like response. Third, because $\Phi/M$ and the Hubble rate redshift differently, an expanding cosmological trajectory moves through the local instability chart. We test these statements by computing the real Floquet exponent from the canonical transverse and longitudinal equations, Eqs.~\eqref{eq:transverseeq} and \eqref{eq:longitudinaleq}. The scale factor is held fixed over each background oscillation, as appropriate for the local Floquet approximation when $H\ll\mphi$. The resulting charts are used only to characterize the local instability structure before cosmological redshifting is applied.

The numerical calculation uses the same canonical variables as the analytic treatment. We obtain the monodromy matrix over one background period with a vectorized Runge--Kutta integration and extract the corresponding Floquet exponent. Growth rates are quoted in units of $\mphi$, and the reference instability contour is defined by $\Re(\mu)/\mphi=0.02$. For each amplitude and detuning, $k_{\rm peak}$ denotes the momentum that maximizes $\Re(\mu)$. To avoid ratios dominated by numerical values in stable regions, the polarization and derivative diagnostics are evaluated only above the adopted instability threshold. The resulting domains are compared with the local half-mass Floquet structure of Ref.~\cite{Adshead:2023}. The monodromy procedure follows standard applications in preheating and vector-resonance calculations \cite{Kofman:1997,Greene:1997,Amin:2014,Dror:2018,Nakai:2023,Barrie:2022qni}. The numerical calculation is local in time and is not used to evolve the cosmological background itself.

At $\mA/\mphi=1/2$, the canonical longitudinal system gives stronger support to the infrared part of the first instability band than the transverse system. The transverse modes instead occupy a wider momentum range and contain a secondary instability tongue at larger momentum. This difference affects the produced spectrum: the transverse sector broadens its momentum support, whereas the longitudinal sector contributes more strongly to its lowest-momentum part.

Comparing the full canonical equations with the approximation that retains only the modulated mass term isolates the derivative contributions from canonical normalization. Their effect is largest inside the transverse instability tongues and remains visible near the edges of the longitudinal domains. These terms therefore affect the band widths, the relative transverse-longitudinal weighting, and the time at which an expanding trajectory first enters a region of efficient local growth.

\begin{figure*}[tbp]
\centering
\includegraphics[width=0.98\textwidth]{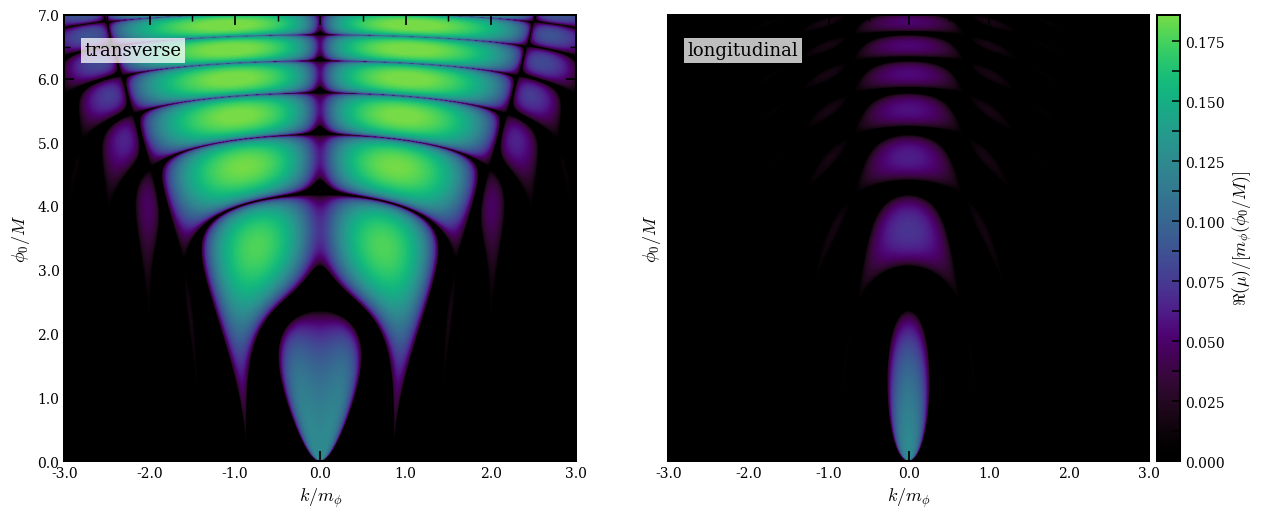}
\caption{Floquet exponents for the tuned branch, $\mA/\mphi = 1/2$, in the transverse and longitudinal sectors as functions of $k/\mphi$ and $\phi_0/M$. The local half-mass structure is consistent with the benchmark calculation of Ref.~\cite{Adshead:2023}; the exponents shown here are obtained from the canonical variables in Eqs.~\eqref{eq:transverseeq} and \eqref{eq:longitudinaleq}. The transverse instability extends over a wider momentum interval and contains a secondary higher-$k$ tongue, while the longitudinal instability is more concentrated toward the infrared.}
\label{fig:bands}
\end{figure*}

\begin{figure*}[tbp]
\centering
\includegraphics[width=0.98\textwidth]{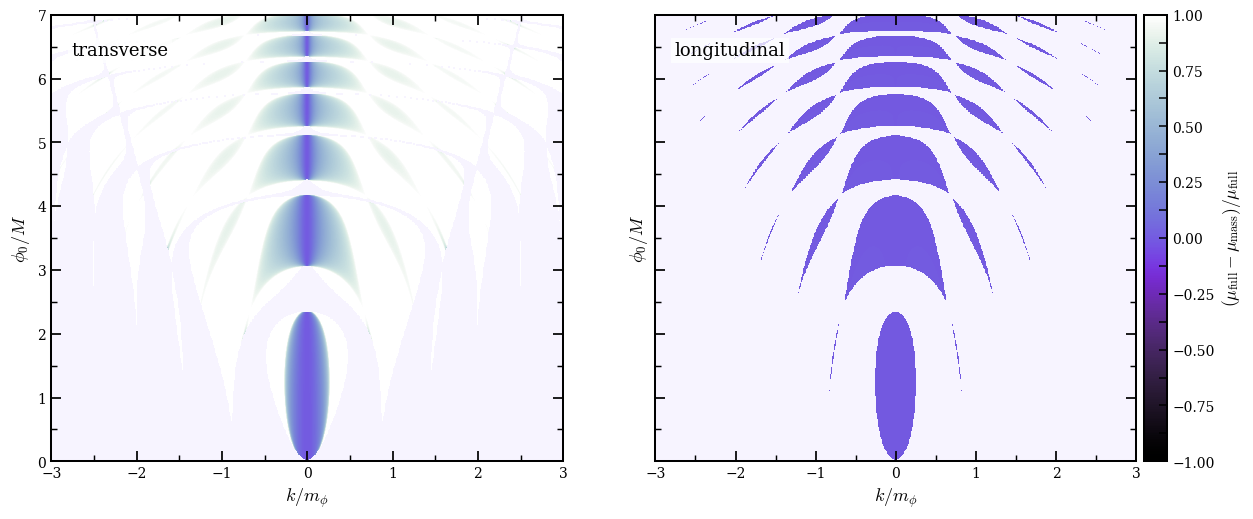}
\caption{Relative contribution of the $\W$-derivative terms to the Floquet exponent, defined by $(\mu_{\rm full}-\mu_{\rm mass})/\mu_{\rm full}$ in the $(k/\mphi,\phi_0/M)$ plane. Left: transverse sector. Right: longitudinal sector. Points below the instability threshold are masked. The derivative terms modify the transverse tongues and shift the boundaries of the longitudinal domains, showing the role of canonical normalization in a polarization-resolved calculation.}
\label{fig:derivimpact}
\end{figure*}

\begin{figure}[tbp]
\centering
\includegraphics[width=0.98\linewidth]{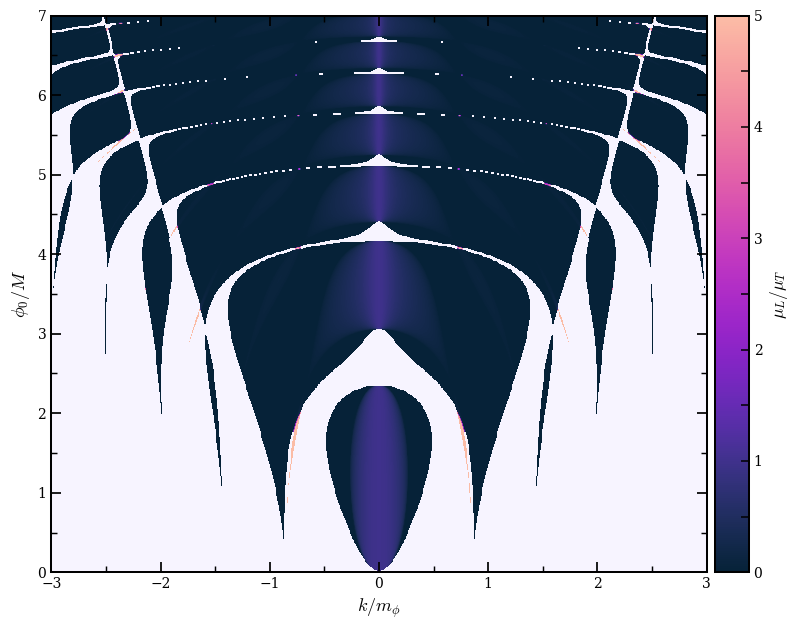}
\caption{Polarization ratio $\mu_L/\mu_T$ over the instability domain. The ratio is shown only where both polarizations are above the reference instability threshold, so divisions involving stable regions are excluded. Values $\mu_L/\mu_T>1$ denote longitudinally dominated growth, whereas $\mu_L/\mu_T<1$ denote transverse-dominated growth.}
\label{fig:polenhance}
\end{figure}

The ratio $\mu_L/\mu_T$ specifies where the longitudinal canonical mode grows faster than the transverse mode. The longitudinal enhancement occupies part of the infrared-centered branch, while the transverse instability extends over a broader momentum interval. Consequently, the polarization that dominates while $\Phi/M$ remains appreciable affects both the spectral distribution and the relic abundance \cite{Dror:2018,Adshead:2023,Kitajima:2023,Nakai:2020}. This distinction is retained when interpreting the momentum support of the final vector population.

\begin{figure}[tbp]
\centering
\includegraphics[width=0.98\linewidth]{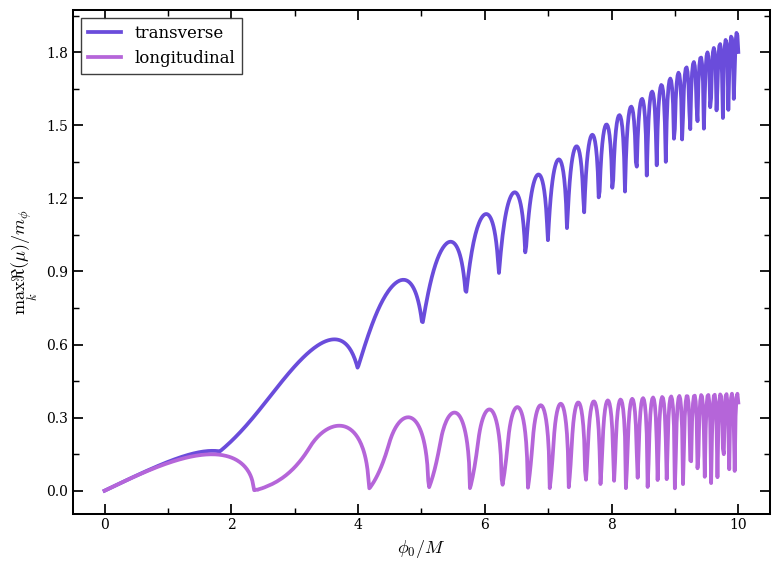}\\[0.4em]
\includegraphics[width=1.0\linewidth]{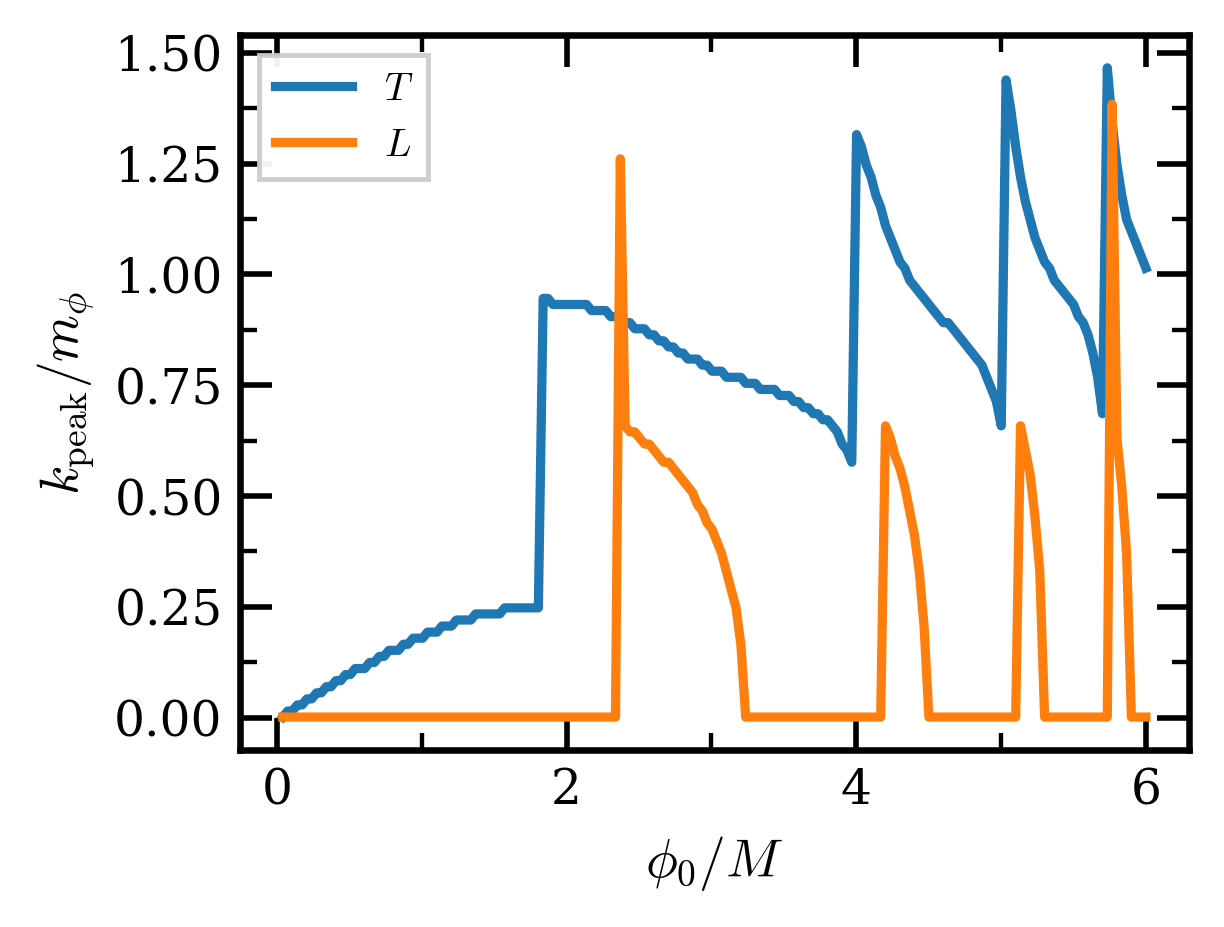}
\caption{Amplitude dependence of the tuned branch. Upper panel: maximum growth rate, $\max_k \Re(\mu_{T,L})/\mphi$, for the transverse and longitudinal sectors. The transverse maximum increases smoothly over the plotted interval, while the longitudinal maximum is nonmonotonic as neighboring longitudinal tongues exchange dominance with increasing $\phi_0/M$. Lower panel: momentum $k_{\rm peak}/\mphi$ at which the Floquet exponent is maximal. The transverse maximum moves toward intermediate momentum, whereas the longitudinal maximum remains near the infrared over most of the perturbative interval and shifts when another tongue becomes dominant.}
\label{fig:ampdiag}
\end{figure}

The amplitude dependence extends the analytic narrow-band discussion beyond the strict small-$\epsilon$ limit. The transverse peak increases smoothly with the modulation amplitude, whereas the longitudinal peak is nonmonotonic because neighboring instability tongues exchange dominance. Over most of the perturbative interval, the longitudinal value of $k_{\rm peak}$ remains closer to the infrared than the transverse one. The longitudinal branch therefore assigns greater weight to the lowest-momentum part of the produced spectrum. The effect is visible before the perturbative approximation completely breaks down.

We characterize departures from the half-mass condition by the detuning parameter
\begin{equation}
 \delta \equiv \left(\frac{\mA}{\mphi}\right)^2 - \frac14.
 \label{eq:delta}
\end{equation}

\begin{figure}[tbp]
\centering
\includegraphics[width=1.0\linewidth]{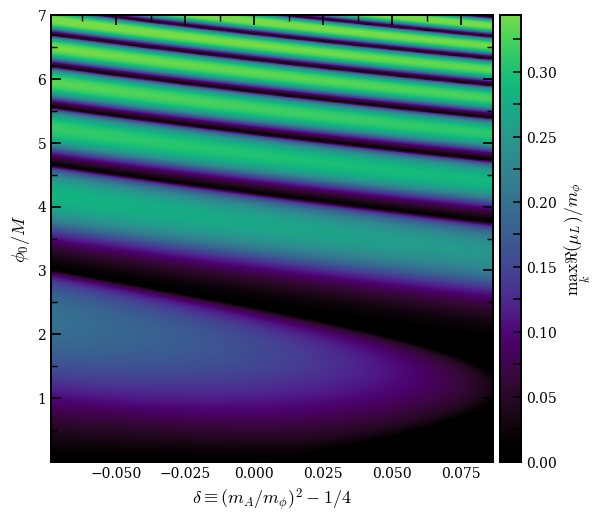}
\caption{Dependence of the tuned branch on detuning. Upper panel: maximum longitudinal growth, $\max_k \Re(\mu_L)/\mphi$, in the $(\phi_0/M,\delta)$ plane, with $\delta \equiv (\mA/\mphi)^2-1/4$.}
\label{fig:detuningmap}
\end{figure}

The detuning map shows a finite interval around $\delta=0$ in which amplification persists, with the largest growth rates close to the exact half-mass condition. Increasing the oscillation amplitude broadens this interval, although the instability remains centered on the tuned branch. The numerical result is consistent with the analytic interpretation of a finite-width, infrared-centered resonance whose growth and momentum support are polarization dependent. During cosmological evolution the system therefore moves through an instability band of finite width rather than remaining at a single resonance point.

\section{Phenomenological implications and benchmark mass ranges}
\label{sec:pheno}

We now combine the cosmological embedding with the polarization-resolved Floquet structure to discuss the relic abundance. The background evolution fixes the spectator history, while the instability analysis specifies the infrared part of the half-mass branch. For the abundance normalization, we use the efficient-transfer, nonrelativistic result for the dilatonic half-mass scenario obtained in Ref.~\cite{Adshead:2023}. The bound in Eq.~\eqref{eq:Mupper} and the scale ordering in Fig.~\ref{fig:trajectory} are independent of this normalization. {Within this phenomenological translation, Ref.~\cite{Adshead:2023} supplies the efficient-transfer normalization. The pre-domination criterion, polarization-resolved canonical dynamics, detuning map, and ultraviolet consistency conditions are the ingredients derived in the preceding sections.} Identifying $m_{\gamma'}\equiv \mA$, the relic density is
\begin{equation}
 \frac{\Omega_{\gamma'} h^2}{0.12}
 \simeq
 \left(\frac{m_{\gamma'}}{10^{-17}\,{\rm eV}}\right)^{1/2}
 \left(\frac{\Phii}{10^{16}\,{\rm GeV}}\right)^2 ,
 \label{eq:abundanceAdshead}
\end{equation}
with the $m_{\gamma'}^{1/2}$ dependence characteristic of the tuned half-mass normalization. The estimate assumes efficient transfer into a narrow vector distribution that becomes nonrelativistic. Polarization splitting changes the detailed momentum support, but within the perturbative narrow-band regime we use Eq.~\eqref{eq:abundanceAdshead} for the total abundance. Substituting Eq.~\eqref{eq:ri} for the initial amplitude gives
\begin{equation}
 \frac{\Omega_{\gamma'} h^2}{0.12}
 \simeq
 3.56\times 10^{5}\, r_i
 \left(\frac{m_{\gamma'}}{10^{-17}\,{\rm eV}}\right)^{1/2},
 \label{eq:abundanceRi}
\end{equation}
and therefore
\begin{equation}
 m_{\gamma'}
 \simeq
 7.9\times 10^{-29}\,{\rm eV}\;
 \left(\frac{\Omega_{\gamma'} h^2}{0.12}\right)^2
 r_i^{-2}.
 \label{eq:massRi}
\end{equation}
The last relation expresses the abundance result directly in terms of the gravitational onset fraction used in the cosmological analysis. This use of the abundance relation is deliberately separated from the preceding derivation: the latter determines when the instability can operate consistently, whereas the normalization converts that allowed production history into a present-day mass and density.

Equation~\eqref{eq:massRi} gives the direct relation between the cosmological initial condition and the relic mass. At fixed abundance, a larger onset fraction corresponds to a smaller vector mass, $m_{\gamma'}\propto r_i^{-2}$. Increasing the onset fraction therefore shifts the relic toward lower masses rather than enlarging the usual ultralight interval. Within this abundance relation, early spectator relevance and the standard ultralight vector mass range prefer different regions of parameter space.

For the representative interval $m_{\gamma'}\sim10^{-20}$--$10^{-18}\,{\rm eV}$ \cite{Caputo:2021,Cyncynates:2025prd}, Eq.~\eqref{eq:massRi} gives
\begin{align}
 r_i(10^{-20}\,{\rm eV}) &\simeq 8.9\times 10^{-5}, \nonumber\\
 r_i(10^{-19}\,{\rm eV}) &\simeq 2.8\times 10^{-5}, \nonumber\\
 r_i(10^{-18}\,{\rm eV}) &\simeq 8.9\times 10^{-6},
 \label{eq:riBench}
\end{align}
so the corresponding spectator is subdominant at the onset of oscillations and the background remains radiation dominated. If instead $r_i={\cal O}(1)$ is imposed in the same abundance relation, the mass is reduced to $m_{\gamma'}\lesssim10^{-28}\,\mathrm{eV}$, below the standard postinflationary ultralight dark-matter interval considered here. Thus the abundance relation selects a clearly subdominant initial condensate.

Equation~\eqref{eq:massRi} thus defines a monotonic curve in the $(r_i,m_{\gamma'})$ plane: at fixed abundance, increasing the onset fraction lowers the relic mass. The ultralight interval of interest lies well below the spectator-domination threshold $r_i\sim1$. Within the transfer approximation, vector production therefore occurs while the dominant postreheating component still controls the expansion. For the benchmark kinetic scale $M=10^{17}\,\mathrm{GeV}$, Eq.~\eqref{eq:exactmap} becomes
\begin{equation}
 r_i \simeq 2.8\times 10^{-4}\,\epsilon_i^2
 \left(\frac{M}{10^{17}\,{\rm GeV}}\right)^2 .
 \label{eq:riMbenchmark}
\end{equation}
Hence $\epsilon_i\lesssim1$ implies $r_i\lesssim3\times10^{-4}$ at this scale. The perturbative half-mass branch is then realized as a delayed radiation-era conversion. Its relative efficiency increases according to $\mu/H\propto a^{1/2}$ instead of relying on an early matter-dominated phase. The same benchmark also illustrates why the small-amplitude condition and spectator subdominance are compatible for a sub-Planckian kinetic scale: both follow simultaneously from the same parameter relation once the kinetic scale is fixed.

\begin{table}[tbp]
\caption{Representative tuned-branch benchmarks for $M=10^{17}\,\mathrm{GeV}$, obtained from Eqs.~\eqref{eq:exactmap} and \eqref{eq:abundanceAdshead}.}
\label{tab:benchmarks}
\begin{ruledtabular}
\begin{tabular}{cccc}
$\epsilon_i=\Phii/M$ & $\Phii\,[{\rm GeV}]$ & $r_i$ & $m_{\gamma'}\,[{\rm eV}]$ \\
\hline
$0.10$ & $1.0\times10^{16}$ & $2.8\times10^{-6}$ & $1.0\times10^{-17}$ \\
$0.32$ & $3.2\times10^{16}$ & $2.8\times10^{-5}$ & $1.0\times10^{-19}$ \\
$1.00$ & $1.0\times10^{17}$ & $2.8\times10^{-4}$ & $1.0\times10^{-21}$ \\
\end{tabular}
\end{ruledtabular}
\end{table}

Table~\ref{tab:benchmarks} gives numerical examples for $M=10^{17}\,\mathrm{GeV}$. Over $\epsilon_i=0.1$--$1$, the corresponding dark-photon mass changes from $10^{-17}$ to $10^{-21}\,\mathrm{eV}$, while the spectator remains below the domination threshold. The intermediate value $\epsilon_i\simeq0.32$ gives $m_{\gamma'}\simeq10^{-19}\,\mathrm{eV}$ and $r_i\simeq2.8\times10^{-5}$, where the abundance relation and the cosmological embedding are simultaneously satisfied within the perturbative branch. If the causal condition $m\gtrsim10^{-19}\,\mathrm{eV}$ applies to this production history, the $\epsilon_i=1$ benchmark lies below the allowed mass, the $\epsilon_i\simeq0.32$ point is near the boundary, and the $\epsilon_i=0.10$ example remains above it. These examples are intended as reference points for the scaling relations rather than as an independent scan of model parameters.

The conversion epoch can be written in terms of the same variables. The tuned-branch estimate of Ref.~\cite{Adshead:2023},
\begin{equation}
 \frac{1+z_\star}{1.9\times 10^5}
 \simeq
 \epsilon_i^2
 \left(\frac{m_{\gamma'}}{10^{-17}\,{\rm eV}}\right)^{1/2},
 \label{eq:zstarAdshead}
\end{equation}
becomes
\begin{equation}
 \frac{1+z_\star}{1.9\times 10^5}
 \simeq
 6\,r_i
 \left(\frac{\Mpl}{M}\right)^2
 \left(\frac{m_{\gamma'}}{10^{-17}\,{\rm eV}}\right)^{1/2}.
 \label{eq:zstarRi}
\end{equation}
This form separates $r_i$ from the kinetic scale $M$. For $M=10^{17}\,\mathrm{GeV}$ and $m_{\gamma'}\simeq10^{-19}\,\mathrm{eV}$, one obtains $1+z_\star\sim2\times10^5$, so the conversion occurs at an early postinflationary epoch. Equations~\eqref{eq:zstarRi} and \eqref{eq:UpsilonScaling} also show that radiation domination increases the Floquet rate relative to Hubble expansion. If the spectator dominates early, the background becomes effectively matter-like and this enhancement is lost. The numerical instability charts are consistent with the same picture: the dominant part of the tuned branch remains centered at low momentum while its relative efficiency increases during radiation domination. In this rewriting, the conversion redshift is not an independent parameter. It is determined by the onset fraction, the kinetic scale, and the relic mass through the same abundance prescription used above.

A further restriction follows if the causal lower bound for postinflationary wave dark matter, $m\gtrsim10^{-19}\,\mathrm{eV}$, is applicable to this mechanism \cite{AminMirbabayi:2024}. Equation~\eqref{eq:massRi} then gives
\begin{equation}
 r_i \lesssim 2.8\times 10^{-5}.
 \label{eq:causalri}
\end{equation}
Under this assumption, the tuned interval is approximately $m_{\gamma'}\sim10^{-19}$--$10^{-18}\,\mathrm{eV}$, corresponding to $r_i\sim10^{-5}$--$10^{-6}$. This region lies in the perturbative radiation-era regime and remains far below the spectator fraction required to control the expansion at the onset of oscillations. The conclusion depends on applying the causal bound to the present production history.

\section{Conclusions}
\label{sec:conclusion}

We have studied the cosmological and ultraviolet conditions for producing ultralight vector dark matter through a dilatonic half-mass resonance sourced by a spectator scalar condensate. The local half-mass branch is an infrared instability selected by the mass ratio, while its cosmological realization depends on three independent quantities: the microscopic modulation $\epsilon_i=\Phii/M$, the gravitational onset fraction $r_i=\Phii^2/(6\Mpl^2)$, and the ultraviolet mechanism that generates the vector mass.

The cosmological condition follows from the competition between the Floquet rate and Hubble expansion. In the linear narrow-band regime, $\mu/H\propto a^{3w_b/2}$. The ratio therefore increases during radiation domination and remains constant for a matter-like background. Requiring efficient amplification to begin before the spectator dominates gives
$
{M}/{\Mpl}\lesssim \sqrt6\,c_1 \simeq 0.31$,
for $c_1\simeq1/8$ at the tuned band center. The expanding-background trajectories give the same ordering: for sub-Planckian kinetic scales, $a_\star<a_{\rm dom}$, whereas for a Planckian scale spectator domination occurs first. This is a condition on the cosmological embedding rather than on the final abundance. Sufficient integrated growth is still required before the scalar oscillation amplitude is diluted by the expansion. The cancellation of the onset amplitude in this ordering is important because it separates the scale bound from the question of whether a particular initial displacement supplies enough total amplification.

Using the efficient-transfer abundance normalization gives $m_{\gamma'}\propto r_i^{-2}$. For
$
10^{-20}\,{\rm eV}\lesssim m_{\gamma'}\lesssim10^{-18}\,{\rm eV}$,
the corresponding onset fractions are $r_i\sim10^{-5}$--$10^{-4}$, so the spectator remains well below the domination threshold during the relevant radiation-dominated epoch. Large initial spectator fractions therefore correspond, within this normalization, to masses below the usual ultralight interval. The polarization-resolved Floquet calculation gives an additional distinction between the two vector sectors. The longitudinal instability is more concentrated toward the infrared, whereas the transverse instability extends over a wider momentum interval and contains secondary tongues. Derivative terms generated by canonical normalization change the Floquet exponents at order unity in parts of the instability domain and must be retained when the two polarizations are compared. The polarization result concerns the shape and weighting of the produced distribution rather than the abundance normalization itself. It therefore supplies information that is not contained in the single relic-density relation.

The ultraviolet requirements depend on the origin of the vector mass. A St\"uckelberg realization does not involve symmetry restoration or topological-defect formation, but the time-dependent kinetic sector and the effective-theory cutoff must remain controlled. A Higgsed realization introduces the radial mode and additional conditions from radial decoupling and possible symmetry restoration caused by the produced vectors. The resonant history must remain within the broken phase if the Proca description is to remain applicable.
{The scale-ordering criterion supplies the broader result of the analysis: for quadratic spectator condensates with a leading Floquet exponent linear in the oscillation amplitude, it converts the local instability rate into a cosmological pre-domination condition, while the dilatonic vector system supplies the explicit polarization-resolved benchmark.}
Taken together, the results identify the conditions under which the dilatonic half-mass branch can be used as a controlled production mechanism for ultralight vector dark matter. The relevant regime has perturbative kinetic modulation, radiation-era evolution, infrared-weighted polarization-dependent production, and an ultraviolet realization consistent with the production scales. The restrictions arise separately from cosmological evolution, local resonance dynamics, and the ultraviolet description of the vector mass. None of these requirements is sufficient alone. The Floquet solution, cosmological ordering, and ultraviolet validity conditions must hold over the production interval.

\begin{acknowledgments}
I.K. acknowledges support from Zhejiang Normal University through a postdoctoral fellowship under Grant No.~YS304224924.
\end{acknowledgments}

\appendix

\section{Derivation of the longitudinal quadratic action}

Equation~\eqref{eq:SL} follows by eliminating the nondynamical temporal component of the massive vector. Starting from the quadratic vector action in Fourier space and projecting onto the longitudinal mode gives
\begin{equation}
\mathcal{L}_{L}^{(2)} =
\frac12 \W \left|A_L' - \ii k A_0\right|^2
\frac12 a^2\mA^2 |A_0|^2
- \frac12 a^2\mA^2 |A_L|^2.
  \end{equation}
Because $A_0$ has no time derivative, it is an auxiliary field fixed algebraically by its equation of motion. Variation with respect to $A_0^*$ gives the constraint in Eq.~\eqref{eq:A0constraint}. Eliminating $A_0$ with that relation yields
  \begin{equation}
  \mathcal{L}_{L}^{(2)} =
  \frac12
  \frac{\W a^2\mA^2}{k^2\W + a^2\mA^2}
  |A_L'|^2
-\frac12 a^2\mA^2 |A_L|^2.
\end{equation}
The resulting kinetic factor depends explicitly on momentum, which is the characteristic feature of the longitudinal sector. With the canonical variable
$
v_L = z_L A_L,
$
and $z_L$ defined in Eq.~\eqref{eq:zLdef}, the action takes canonical form and gives the oscillator equation in Eqs.~\eqref{eq:SL}--\eqref{eq:longitudinaleq}. This derivation shows directly why the longitudinal normalization differs from the transverse one and why its derivative term must be kept in the polarization-resolved Floquet analysis. The auxiliary-field elimination is therefore necessary before assigning a canonical Floquet frequency to the physical longitudinal polarization.

\section{Notation and numerical conventions}

We collect here the notation and numerical conventions used in the calculation. The symbol $\Phi$ denotes the time-dependent envelope of the spectator oscillation, and $\Phii$ is its value when coherent oscillations begin. The resonance parameter is
$
\epsilon(t)\equiv {\Phi(t)}/{M},
$
with onset value $\epsi\equiv \Phii/M$. The initial spectator fraction $r_i$ is evaluated at $H=\mphi$, and $w_b$ is the equation-of-state parameter of the dominant cosmological background.

The instability charts are obtained from the canonical transverse and longitudinal equations, Eqs.~\eqref{eq:transverseeq} and \eqref{eq:longitudinaleq}, using the local Floquet approximation in which the scale factor is fixed during one background oscillation. The monodromy matrix is calculated by evolving two independent initial-condition vectors over one period,
$
0\le m_\phi t\le 2\pi,$
with a fourth-order Runge--Kutta integrator. Selected parameter points were also evaluated with a DOP853 monodromy extraction. Both procedures give the same small-amplitude coefficient,
$
c_1=\max_k\left({\mu}/{\epsilon m_\phi}\right)\simeq0.125,
$
at $\mA/\mphi=1/2$.
For the main instability maps we use
$
\phi_0/M\in[0.04,6],
\quad
k/m_\phi\in[0.001,3],
$
and for the detuning scan
$
m_A/m_\phi\in[0.42,0.58].
$
Growth rates below the plotting floor are set to zero. In the polarization-ratio and derivative diagnostics, points below the adopted instability threshold are masked to avoid divisions by small values in stable regions. The contour boundaries in the instability plots correspond to
$
\Re(\mu)/m_\phi = 0.02.
$
These choices are applied uniformly to the transverse and longitudinal calculations over the parameter ranges shown. The numerical threshold is used only to display and compare the instability domains; it is not inserted into the analytic derivation of the cosmological scale-ordering condition.

\bibliographystyle{apsrev4-2}
\bibliography{references}

@article{Hui:2017,
  author = {Hui, Lam and Ostriker, Jeremiah P. and Tremaine, Scott and Witten, Edward},
  title = {{Ultralight scalars as cosmological dark matter}},
  journal = {Phys. Rev. D},
  volume = {95},
  number = {4},
  pages = {043541},
  year = {2017},
  doi = {10.1103/PhysRevD.95.043541},
  eprint = {1610.08297},
  archivePrefix = {arXiv},
  primaryClass = {astro-ph.CO}
}

@article{Graham:2015,
  author = {Graham, Peter W. and Mardon, Jeremy and Rajendran, Surjeet},
  title = {{Vector Dark Matter from Inflationary Fluctuations}},
  journal = {Phys. Rev. D},
  volume = {93},
  number = {10},
  pages = {103520},
  year = {2016},
  doi = {10.1103/PhysRevD.93.103520},
  eprint = {1504.02102},
  archivePrefix = {arXiv},
  primaryClass = {hep-ph}
}

@article{Khan:2026nsz,
  author = {Khan, Imtiaz and Pirzada, Mustafa G.},
  title = {Post-Inflationary Quenched Production of Axion SU(2) Dark Matter},
  journal = {arXiv e-prints},
  year = {2026},
  eprint = {2604.07044},
  archivePrefix = {arXiv},
  primaryClass = {hep-ph},
  month = {4}
}

@article{Dror:2018,
  author = {Dror, Jeff A. and Harigaya, Keisuke and Narayan, Vijay},
  title = {{Parametric Resonance Production of Ultralight Vector Dark Matter}},
  journal = {Phys. Rev. D},
  volume = {99},
  number = {3},
  pages = {035036},
  year = {2019},
  doi = {10.1103/PhysRevD.99.035036},
  eprint = {1810.07195},
  archivePrefix = {arXiv},
  primaryClass = {hep-ph}
}

@article{Co:2018,
  author = {Co, Raymond T. and Pierce, Aaron and Zhang, Zhengkang and Zhao, Yue},
  title = {{Dark Photon Dark Matter Produced by Axion Oscillations}},
  journal = {Phys. Rev. D},
  volume = {99},
  number = {7},
  pages = {075002},
  year = {2019},
  doi = {10.1103/PhysRevD.99.075002},
  eprint = {1810.07196},
  archivePrefix = {arXiv},
  primaryClass = {hep-ph}
}

@article{Agrawal:2018,
  author = {Agrawal, Prateek and Kitajima, Naoya and Reece, Matthew and Sekiguchi, Toyokazu and Takahashi, Fuminobu},
  title = {{Relic Abundance of Dark Photon Dark Matter}},
  journal = {Phys. Lett. B},
  volume = {801},
  pages = {135136},
  year = {2020},
  doi = {10.1016/j.physletb.2019.135136},
  eprint = {1810.07188},
  archivePrefix = {arXiv},
  primaryClass = {hep-ph}
}

@article{Long:2019,
  author = {Long, Andrew J. and Wang, Lian-Tao},
  title = {{Dark Photon Dark Matter from a Network of Cosmic Strings}},
  journal = {Phys. Rev. D},
  volume = {99},
  number = {6},
  pages = {063529},
  year = {2019},
  doi = {10.1103/PhysRevD.99.063529},
  eprint = {1901.03312},
  archivePrefix = {arXiv},
  primaryClass = {hep-ph}
}

@article{Kofman:1997,
  author = {Kofman, Lev and Linde, Andrei D. and Starobinsky, Alexei A.},
  title = {{Towards the theory of reheating after inflation}},
  journal = {Phys. Rev. D},
  volume = {56},
  pages = {3258--3295},
  year = {1997},
  doi = {10.1103/PhysRevD.56.3258},
  eprint = {hep-ph/9704452},
  archivePrefix = {arXiv}
}

@article{Amin:2014,
  author = {Amin, Mustafa A. and Hertzberg, Mark P. and Kaiser, David I. and Karouby, Johanna},
  title = {{Nonperturbative Dynamics Of Reheating After Inflation: A Review}},
  journal = {Int. J. Mod. Phys. D},
  volume = {24},
  pages = {1530003},
  year = {2015},
  doi = {10.1142/S0218271815300037},
  eprint = {1410.3808},
  archivePrefix = {arXiv},
  primaryClass = {hep-ph}
}

@article{Adshead:2023,
  author = {Adshead, Peter and Lozanov, Kaloian D. and Weiner, Zachary J.},
  title = {{Dark photon dark matter from an oscillating dilaton}},
  journal = {Phys. Rev. D},
  volume = {107},
  number = {8},
  pages = {083519},
  year = {2023},
  doi = {10.1103/PhysRevD.107.083519},
  eprint = {2301.07718},
  archivePrefix = {arXiv},
  primaryClass = {hep-ph}
}

@article{Kitajima:2023,
  author = {Kitajima, Naoya and Takahashi, Fuminobu},
  title = {{Resonant production of dark photons from axions without a large coupling}},
  journal = {Phys. Rev. D},
  volume = {107},
  number = {12},
  pages = {123518},
  year = {2023},
  doi = {10.1103/PhysRevD.107.123518},
  eprint = {2303.05492},
  archivePrefix = {arXiv},
  primaryClass = {hep-ph}
}

@article{Kitajima:2022strings,
  author = {Kitajima, Naoya and Nakayama, Kazunori},
  title = {{Dark photon dark matter from cosmic strings and gravitational wave background}},
  journal = {JHEP},
  volume = {2023},
  number = {08},
  pages = {068},
  year = {2023},
  doi = {10.1007/JHEP08(2023)068},
  eprint = {2212.13573},
  archivePrefix = {arXiv},
  primaryClass = {hep-ph}
}

@misc{Cline:2024,
  author = {Cline, James M.},
  title = {{Vector dark matter, inflation, and non-minimal couplings}},
  archivePrefix = {arXiv},
  primaryClass = {hep-ph},
  year = {2024},
  note = {INSPIRE-listed review/analysis of vector dark-matter production and nonminimal couplings}
}

@article{Cyncynates:2025prl,
  author = {Cyncynates, David and Weiner, Zachary J.},
  title = {{Detectable and Defect-Free Dark Photon Dark Matter}},
  journal = {Phys. Rev. Lett.},
  volume = {134},
  number = {21},
  pages = {211002},
  year = {2025},
  doi = {10.1103/PhysRevLett.134.211002},
  eprint = {2310.18397},
  archivePrefix = {arXiv},
  primaryClass = {hep-ph}
}

@article{Cyncynates:2025prd,
  author = {Cyncynates, David and Weiner, Zachary J.},
  title = {{Experimental targets for dark photon dark matter}},
  journal = {Phys. Rev. D},
  volume = {111},
  number = {10},
  pages = {103535},
  year = {2025},
  doi = {10.1103/PhysRevD.111.103535},
  eprint = {2410.14774},
  archivePrefix = {arXiv},
  primaryClass = {hep-ph}
}

@article{Kitajima:2025bound,
  author = {Kitajima, Naoya and Nakagawa, Shota and Takahashi, Fuminobu and Yin, Wen},
  title = {{A bound on light dark photon dark matter}},
  journal = {Phys. Lett. B},
  volume = {862},
  pages = {139304},
  year = {2025},
  doi = {10.1016/j.physletb.2025.139304},
  eprint = {2410.17964},
  archivePrefix = {arXiv},
  primaryClass = {hep-ph}
}

@article{Pirzada_2026uak,
  author       = {Pirzada and Khan, Imtiaz and Khan, Mussawair and Li, Tianjun and Muhammad, Ali},
  title        = {Non-Minimal Dilaton Inflation from the Effective Gluodynamics},
  journal      = {arXiv e-prints},
  year         = {2026},
  month        = {3},
  eprint       = {2603.00818},
  archivePrefix= {arXiv},
  primaryClass = {hep-ph}
}

@article{Pirzada:2026npl,
  author       = {Pirzada and Gao, Yu and Yang, Qiaoli},
  title        = {Parametric-Resonance Production of QCD Axions},
  journal      = {arXiv e-prints},
  year         = {2026},
  month        = {2},
  eprint       = {2602.06922},
  archivePrefix= {arXiv},
  primaryClass = {hep-ph}
}

@article{Ijaz:2024zma,
    author = {Ijaz, Nadir and Rehman, Mansoor Ur},
    title = {Exploring primordial black holes and gravitational waves with R-symmetric GUT Higgs inflation},
    eprint = {2402.13924},
    archivePrefix = {arXiv},
    primaryClass = {astro-ph.CO},
    doi = {10.1016/j.physletb.2024.139229},
    journal = {Phys. Lett. B},
    volume = {861},
    pages = {139229},
    year = {2025}
}

@article{Ijaz:2023cvc,
    author = {Ijaz, Nadir and Mehmood, Maria and Rehman, Mansoor Ur},
    title = {The stochastic gravitational-wave background from primordial black holes and observable proton decay in R-symmetric SU(5) Inflation},
    eprint = {2308.14908},
    archivePrefix = {arXiv},
    primaryClass = {astro-ph.CO},
    doi = {10.1140/epjc/s10052-025-15078-w},
    journal = {Eur. Phys. J. C},
    volume = {85},
    number = {12},
    pages = {1394},
    year = {2025}
}

@article{Barbon:2025wjl,
  author       = {Barbon, Matteo and Ijaz, Nadir and Peloso, Marco},
  title        = {Axion inflation in the regime of homogeneous backreaction},
  journal      = {arXiv e-prints},
  year         = {2025},
  month        = {10},
  eprint       = {2510.17207},
  archivePrefix= {arXiv},
  primaryClass = {astro-ph.CO}
}

@article{Khan:2025ibo,
  author       = {Khan, Imtiaz and Muhammad, Ali and Li, Tianjun and Raza, Shabbar and Pirzada and Khan, Mussawir},
  title        = {The Light Neutralino Dark Matter at Future Colliders in the MSSM with the Generalized Minimal Supergravity (GmSUGRA)},
  journal      = {arXiv e-prints},
  year         = {2025},
  month        = {9},
  eprint       = {2509.23356},
  archivePrefix= {arXiv},
  primaryClass = {hep-ph}
}

@article{Muhammad:2026gmg,
  author       = {Muhammad, Ali and Khan, Imtiaz and Li, Tianjun and Raza, Shabbar and Khan, Mussawir and Pirzada},
  title        = {LHC Run-3, Dark Matter and Supersymmetric Spectra in the Supersymmetric Pati-Salam Model},
  journal      = {arXiv e-prints},
  year         = {2026},
  month        = {3},
  eprint       = {2603.24152},
  archivePrefix= {arXiv},
  primaryClass = {hep-ph}
}

@article{AminMirbabayi:2024,
  author = {Amin, Mustafa A. and Mirbabayi, Mehrdad},
  title = {{A Lower Bound on Dark Matter Mass}},
  journal = {Phys. Rev. Lett.},
  volume = {132},
  number = {22},
  pages = {221004},
  year = {2024},
  doi = {10.1103/PhysRevLett.132.221004},
  eprint = {2211.09775},
  archivePrefix = {arXiv},
  primaryClass = {hep-ph}
}

@misc{Nakai:2023,
  author = {Nakai, Yuichiro and others},
  title = {{Resonant Vector Dark Matter Production during Inflation}},
  archivePrefix = {arXiv},
  primaryClass = {hep-ph},
  year = {2023},
  note = {INSPIRE record: Resonant Vector Dark Matter Production during Inflation}
}

@article{Holdom:1986,
  author = {Holdom, Bob},
  title = {{Two U(1)'s and Epsilon Charge Shifts}},
  journal = {Phys. Lett. B},
  volume = {166},
  pages = {196--198},
  year = {1986},
  doi = {10.1016/0370-2693(86)91377-8}
}

@article{Nelson:2011,
  author = {Nelson, Ann E. and Scholtz, Jakub},
  title = {{Dark Light, Dark Matter and the Misalignment Mechanism}},
  journal = {Phys. Rev. D},
  volume = {84},
  pages = {103501},
  year = {2011},
  doi = {10.1103/PhysRevD.84.103501},
  eprint = {1105.2812},
  archivePrefix = {arXiv},
  primaryClass = {hep-ph}
}

@article{Arias:2012,
  author = {Arias, Paola and Cadamuro, Davide and Goodsell, Mark and Jaeckel, Joerg and Redondo, Javier and Ringwald, Andreas},
  title = {{WISPy Cold Dark Matter}},
  journal = {JCAP},
  volume = {2012},
  number = {06},
  pages = {013},
  year = {2012},
  doi = {10.1088/1475-7516/2012/06/013},
  eprint = {1201.5902},
  archivePrefix = {arXiv},
  primaryClass = {hep-ph}
}

@article{Mirizzi:2009,
  author = {Mirizzi, Alessandro and Redondo, Javier and Sigl, Guenter},
  title = {{Microwave Background Constraints on Mixing of Photons with Hidden Photons}},
  journal = {JCAP},
  volume = {2009},
  number = {03},
  pages = {026},
  year = {2009},
  doi = {10.1088/1475-7516/2009/03/026},
  eprint = {0901.0014},
  archivePrefix = {arXiv},
  primaryClass = {hep-ph}
}

@article{Traschen:1990,
  author = {Traschen, Jennie H. and Brandenberger, Robert H.},
  title = {{Particle Production During Out-of-equilibrium Phase Transitions}},
  journal = {Phys. Rev. D},
  volume = {42},
  pages = {2491--2504},
  year = {1990},
  doi = {10.1103/PhysRevD.42.2491}
}

@article{Shtanov:1995,
  author = {Shtanov, Yuri and Traschen, Jennie H. and Brandenberger, Robert H.},
  title = {{Universe reheating after inflation}},
  journal = {Phys. Rev. D},
  volume = {51},
  pages = {5438--5455},
  year = {1995},
  doi = {10.1103/PhysRevD.51.5438},
  eprint = {hep-ph/9407247},
  archivePrefix = {arXiv}
}

@book{Fabbrichesi:2020,
  author = {Fabbrichesi, Marco and Gabrielli, Emidio and Lanfranchi, Gaia},
  title = {{The Physics of the Dark Photon: A Primer}},
  publisher = {Springer},
  year = {2021},
  doi = {10.1007/978-3-030-62519-1},
  eprint = {2005.01515},
  archivePrefix = {arXiv},
  primaryClass = {hep-ph}
}

@article{Capanelli:2024runaway,
  author = {Capanelli, Christian and Jenks, Leah and Kolb, Edward W. and McDonough, Evan},
  title = {Runaway Gravitational Production of Dark Photons},
  journal = {Phys. Rev. Lett.},
  volume = {133},
  pages = {061602},
  year = {2024},
  doi = {10.1103/PhysRevLett.133.061602},
  eprint = {2403.15536},
  archivePrefix = {arXiv},
  primaryClass = {hep-th}
}

@article{Capanelli:2024gpp,
  author = {Capanelli, Christian and Jenks, Leah and Kolb, Edward W. and McDonough, Evan},
  title = {Gravitational Production of Completely Dark Photons with Nonminimal Couplings to Gravity},
  journal = {JHEP},
  number = {09},
  pages = {071},
  year = {2024},
  doi = {10.1007/JHEP09(2024)071},
  eprint = {2405.19390},
  archivePrefix = {arXiv},
  primaryClass = {hep-th}
}

@article{Caputo:2021,
  author = {Caputo, Andrea and O'Hare, Ciaran A. J. and Millar, Alexander J. and Vitagliano, Edoardo},
  title = {{Dark photon limits: A handbook}},
  journal = {Phys. Rev. D},
  volume = {104},
  number = {9},
  pages = {095029},
  year = {2021},
  doi = {10.1103/PhysRevD.104.095029},
  eprint = {2105.04565},
  archivePrefix = {arXiv},
  primaryClass = {hep-ph}
}

@article{Greene:1997,
  author = {Greene, Patrick B. and Kofman, Lev and Linde, Andrei D. and Starobinsky, Alexei A.},
  title = {{Structure of resonance in preheating after inflation}},
  journal = {Phys. Rev. D},
  volume = {56},
  pages = {6175--6192},
  year = {1997},
  doi = {10.1103/PhysRevD.56.6175},
  eprint = {hep-ph/9705347},
  archivePrefix = {arXiv}
}

@misc{Lozanov:2018,
  author = {Lozanov, Kaloian D.},
  title = {{Lectures on Reheating after Inflation}},
  eprint = {1907.04402},
  archivePrefix = {arXiv},
  primaryClass = {astro-ph.CO},
  year = {2019}
}

@article{Allahverdi:2010,
  author = {Allahverdi, Rouzbeh and Brandenberger, Robert and Cyr-Racine, Francis-Yan and Mazumdar, Anupam},
  title = {{Reheating in Inflationary Cosmology: Theory and Applications}},
  journal = {Ann. Rev. Nucl. Part. Sci.},
  volume = {60},
  pages = {27--51},
  year = {2010},
  doi = {10.1146/annurev.nucl.012809.104511},
  eprint = {1001.2600},
  archivePrefix = {arXiv},
  primaryClass = {hep-th}
}

@article{Nakai:2020,
  author = {Nakai, Yuichiro and Namba, Ryo and Wang, Ziwei},
  title = {{Light Dark Photon Dark Matter from Inflation}},
  journal = {JHEP},
  volume = {2020},
  number = {12},
  pages = {170},
  year = {2020},
  doi = {10.1007/JHEP12(2020)170},
  eprint = {2004.10743},
  archivePrefix = {arXiv},
  primaryClass = {hep-ph}
}

@article{Ozsoy:2023,
  author = {\"Ozsoy, Ogan and Tasinato, Gianmassimo},
  title = {{Vector dark matter, inflation and non-minimal couplings with gravity}},
  journal = {JCAP},
  volume = {2024},
  number = {06},
  pages = {003},
  year = {2024},
  doi = {10.1088/1475-7516/2024/06/003},
  eprint = {2310.03862},
  archivePrefix = {arXiv},
  primaryClass = {hep-th}
}

@article{Capanelli:2024,
  author = {Capanelli, Clelia and Jenks, Leah and Kolb, Edward W. and McDonough, Evan},
  title = {{Gravitational production of completely dark photons with nonminimal couplings to gravity}},
  journal = {JHEP},
  volume = {2024},
  number = {09},
  pages = {071},
  year = {2024},
  doi = {10.1007/JHEP09(2024)071},
  eprint = {2405.19390},
  archivePrefix = {arXiv},
  primaryClass = {hep-ph}
}

@article{Pirzada:2026sle,
  author = {Pirzada and Muhammad, Ali and Li, Tianjun and Khan, Imtiaz and Khan, Mussawir},
  title = {{Dilaton-Flattened Axion Inflation}},
  journal = {arXiv e-prints},
  eprint = {2604.15194},
  archivePrefix = {arXiv},
  primaryClass = {hep-ph},
  month = {4},
  year = {2026}
}

@article{Pirzada:2026jml,
  author = {Pirzada and Li, Tianjun},
  title = {{Controlled Penumbral Inflation from Monodromic Valleys}},
  journal = {arXiv e-prints},
  eprint = {2605.10197},
  archivePrefix = {arXiv},
  primaryClass = {hep-ph},
  month = {5},
  year = {2026}
}

@article{AlonsoAlvarez:2019cgw,
  author = {Alonso-{\'A}lvarez, Gonzalo and Hugle, Thomas and Jaeckel, Joerg},
  title = {{Misalignment \& Co.: (Pseudo-)scalar and vector dark matter with curvature couplings}},
  journal = {JCAP},
  volume = {2020},
  number = {02},
  pages = {014},
  year = {2020},
  doi = {10.1088/1475-7516/2020/02/014},
  eprint = {1905.09836},
  archivePrefix = {arXiv},
  primaryClass = {hep-ph}
}

@article{Ema:2019yrd,
  author = {Ema, Yohei and Nakayama, Kazunori and Tang, Yong},
  title = {{Production of purely gravitational dark matter: the case of fermion and vector boson}},
  journal = {JHEP},
  volume = {2019},
  number = {07},
  pages = {060},
  year = {2019},
  doi = {10.1007/JHEP07(2019)060},
  eprint = {1903.10973},
  archivePrefix = {arXiv},
  primaryClass = {hep-ph}
}

@article{Ahmed:2020fhc,
  author = {Ahmed, Aqeel and Grzadkowski, Bohdan and Socha, Anna},
  title = {{Gravitational production of vector dark matter}},
  journal = {JHEP},
  volume = {2020},
  number = {08},
  pages = {059},
  year = {2020},
  doi = {10.1007/JHEP08(2020)059},
  eprint = {2005.01766},
  archivePrefix = {arXiv},
  primaryClass = {hep-ph}
}

@article{Kolb:2020fwh,
  author = {Kolb, Edward W. and Long, Andrew J.},
  title = {{Completely dark photons from gravitational particle production during inflation}},
  journal = {JHEP},
  volume = {2021},
  number = {03},
  pages = {283},
  year = {2021},
  doi = {10.1007/JHEP03(2021)283},
  eprint = {2009.03828},
  archivePrefix = {arXiv},
  primaryClass = {astro-ph.CO}
}

@article{Barrie:2022qni,
  author = {Barrie, Neil D.},
  title = {{Resonant Vector Dark Matter Production during Inflation}},
  journal = {arXiv e-prints},
  year = {2022},
  eprint = {2211.03902},
  archivePrefix = {arXiv},
  primaryClass = {hep-ph}
}

@article{Co:2021rhi,
  author = {Co, Raymond T. and Harigaya, Keisuke and Pierce, Aaron},
  title = {{Gravitational Waves and Dark Photon Dark Matter from Axion Rotations}},
  journal = {arXiv e-prints},
  year = {2021},
  eprint = {2104.02077},
  archivePrefix = {arXiv},
  primaryClass = {hep-ph}
}

@article{Redi:2022zkt,
  author = {Redi, Michele and Tesi, Andrea},
  title = {{Dark photon Dark Matter without Stueckelberg mass}},
  journal = {JHEP},
  volume = {2022},
  number = {10},
  pages = {167},
  year = {2022},
  doi = {10.1007/JHEP10(2022)167},
  eprint = {2204.14274},
  archivePrefix = {arXiv},
  primaryClass = {hep-ph}
}

@article{Sato:2022cwt,
  author = {Sato, Takanori and Takahashi, Fuminobu and Yamada, Masaki},
  title = {{Gravitational production of dark photon dark matter with mass generated by the Higgs mechanism}},
  journal = {JCAP},
  volume = {2022},
  number = {08},
  pages = {022},
  year = {2022},
  doi = {10.1088/1475-7516/2022/08/022},
  eprint = {2204.11896},
  archivePrefix = {arXiv},
  primaryClass = {hep-ph}
}

@article{Salehian:2020jws,
  author = {Salehian, Borna and Gorji, Mohammad Ali and Firouzjahi, Hassan and Mukohyama, Shinji},
  title = {{Vector dark matter production from inflation with symmetry breaking}},
  journal = {Phys. Rev. D},
  volume = {103},
  pages = {063526},
  year = {2021},
  doi = {10.1103/PhysRevD.103.063526},
  eprint = {2010.04491},
  archivePrefix = {arXiv},
  primaryClass = {hep-ph}
}

@article{Cembranos:2023qph,
  author = {Cembranos, Jose A. R. and Garay, Luis J. and Parra-Lopez, Alvaro and Sanchez Velazquez, Jose M.},
  title = {{Vector dark matter production during inflation and reheating}},
  journal = {JCAP},
  volume = {2024},
  number = {02},
  pages = {013},
  year = {2024},
  doi = {10.1088/1475-7516/2024/02/013},
  eprint = {2310.07515},
  archivePrefix = {arXiv},
  primaryClass = {hep-ph}
}

\end{document}